\documentclass[AMA,LATO2COL]{WileyNJDv5}

\usepackage{soul}
\setstcolor{red}
\usepackage{physics}
\usepackage{multirow,makecell}
\usepackage{amssymb, amsmath} 
\usepackage{bm}

\newcommand{\R}{\mathbb{R}}
\newcommand{\fn}[2]{\mathinner{#1\mathopen{\left(#2\right)}}}
\newcommand{\vect}[1]{{\bf #1}}%{\bm{#1}}
\newcommand{\F}[2]{\fn{F^\mathrm{(#1D)}}{#2}}

\newcommand{\BETA}[2]{\beta^\mathrm{(#1D)}_{#2}}
\newcommand{\tens}[1]{\boldsymbol{#1}}

\newcommand{\ATM}[2]{\fn{A_{#1}^{TM}}{#2}}
\newcommand{\ATE}[2]{\fn{A_{#1}^{TE}}{#2}}
\newcommand{\ETM}[1]{\fn{\varepsilon_\mathrm{e}^\mathrm{TM}}{#1}}
\newcommand{\ETE}[1]{\fn{\varepsilon_\mathrm{e}^\mathrm{TE}}{#1}}
\newcommand{\uvect}[1]{\hat{\vect{#1}}}

\newcommand{\SI}{\textit{Supporting Information}}

\articletype{Research Article}%}%

\received{15 October 2025}
\revised{9 March 2026}
\accepted{31 March 2026}
\journal{Advanced Optical Materials}
\volume{0}
\copyyear{2026}
\artid{e03370}
\doi{10.1002/adom.202503370}
\startpage{1}

\begin{document}

\title{Predictive Formulas for Scattering Mean Free Path for General Disordered Dielectric Media Beyond the Long-Wavelength Regime}

\author[1,2,3,4,5]{Jaeuk Kim}% ORCID: https://orcid.org/0000-0002-5562-2937
\author[1,2,3,6]{Salvatore Torquato}% ORCID: https://orcid.org/0000-0003-4614-335X

\authormark{Kim and Torquato}
\titlemark{Predictive Formulas for Scattering Mean Free Path for General Disordered Dielectric Media Beyond the Long-Wavelength Regime}

\address[1]{\orgdiv{Princeton Materials Institute}, \orgname{Princeton University}, \orgaddress{Princeton, \state{New Jersey}, \country{USA}}}

\address[2]{\orgdiv{Department of Chemistry}, \orgname{Princeton University}, \orgaddress{Princeton, \state{New Jersey}, \country{USA}}}

\address[3]{\orgdiv{Department of Physics}, \orgname{Princeton University}, \orgaddress{Princeton, \state{New Jersey}, \country{USA}}}

\address[4]{\orgdiv{Department of Chemical and Biomolecular Engineering}, \orgname{Korea Advanced Institute of Science and Technology (KAIST)}, \orgaddress{Daejeon, \country{Republic of Korea}}}

\address[5]{\orgdiv{GIST InnoCORE AI-Nano Convergence Institute for Early Detection of Neurodegenerative Diseases}, \orgname{Gwangju Institute of Science and Technology}, \orgaddress{Gwangju, \country{Republic of Korea}}}

\address[6]{\orgdiv{Program in Applied and Computational Mathematics}, \orgname{Princeton University}, \orgaddress{Princeton, \state{New Jersey}, \country{USA}}}

\corres{Salvatore Torquato \email{torquato@princeton.edu}}

%\fundingInfo{Text}
%\JELinfo{ejlje}

\abstract[Abstract]{
The majority of previous formulas for the scattering mean free path $\ell_s$ were derived to treat ordinary disordered particulate media consisting of identical dielectric particles embedded in a matrix of another dielectric constant, including the well-known Mie estimate.
We derive new approximate formulas for $\ell_{s}$ that apply to more general particulate media (e.g., arbitrarily shaped particles of different sizes) and non-particulate media in $d$ dimensions that accurately account for the microstructure via the spectral density.
These approximations are based on the exact strong-contrast expansion for the effective dynamic dielectric constant [Torquato and Kim, Phys. Rev. X 11, 021002 (2021)]. To validate the versatility and accuracy of these new formulas for $\ell_s$, we apply them to five different model microstructures in two and three dimensions, including nonhyperuniform and hyperuniform particulate and non-particulate media.
Using finite-difference time-domain (FDTD) simulations at selected volume fractions, we demonstrate that our predictive formulas for the scattering mean free path $\ell_s$ are accurate beyond the long-wavelength regime, that is, for $k_1/s \lesssim 1$, where $k_1$ is the incident wavenumber, and $s$ denotes the specific surface of the microstructure.
In this regime, the strong-contrast formulas are shown to be consistent with the predictions from the Mie theory but can be notably more accurate for 2D TM polarization modes.
For the specific case of monodisperse sphere packings, this condition ($k_1/s \lesssim 1$) corresponds to a particle diameter-to-wavelength ratio of approximately 0.5 or smaller.
However, the Mie estimates become more accurate for $k_1/s \gtrsim 1$.
For hyperuniform media with a power-law spectral density (i.e., $\tilde{\chi}_{_{V}}(k)\sim k^\alpha$ for small $k$), our formulas predict a scaling behavior $\ell_{s} \sim {k_{1}}^{-(d+1+\alpha)}$.
We also provide corresponding scalings laws for the other 2D and 3D nonhyperuniform and hyperuniform models considered here.
Our work enables the inverse design of novel wave characteristics of disordered hyperuniform and nonhyperuniform media by engineering their spectral densities.

}


\maketitle

\renewcommand\thefootnote{}
\footnotetext{\textbf{Abbreviations:} SHU, Stealthy hyperuniform; FDTD simulation, finite-difference time-domain simulation;  TE, Transverse electric; TM, transverse magnetic}

\renewcommand\thefootnote{\fnsymbol{footnote}}
\setcounter{footnote}{1}

\section{Introduction}\label{sec:intro}

For electromagnetic waves propagating in complex two-phase
dielectric media, the scattering mean free path $\ell_s$ is a fundamental length scale that quantifies the characteristic decay length of the coherent wave due to scattering, and physically represents the average distance traveled by light between successive scattering events.\cite{sheng_introduction_2006,akkermans_mesoscopic_2007,vynck_light_2023}
This length scale, or equivalently the scattering coefficient $\mu_s\equiv1/\ell_s$, provides estimates of
wave energy flux attenuation,\cite{monsarrat_pseudogap_2022,
vynck_light_2023}, transparency threshold,\cite{leseur_highdensity_2016} and classifications of distinct transport regimes.\cite{zhang_wave_1999,ioffe_noncrystalline_1960,
wiersma_disordered_2013}
For a macroscopic sample length $L$, three regimes are commonly specified by the magnitude of the ratio $L/\ell_s$, namely, ballistic
($L/\ell_s \ll 1$), single-scattering ($L/\ell_s \sim 1$), and diffusive or multiple-scattering ($L/\ell_s \gg 1$) regimes.\cite{zhang_wave_1999}
Additional regimes can be characterized by the dimensionless quantity $k_\mathrm{e} \ell_s$, which measures the strength of scattering, where the effective wavenumber of this medium $k_\mathrm{e}=\sqrt{\varepsilon_\mathrm{e}}k_0$ is directly determined by the effective dielectric constant $\varepsilon_\mathrm{e}$ and the vacuum wavenumber $k_0$.
In particular, the weak-scattering regime corresponds to the condition $k_\mathrm{e} \ell_s \gg 1$,\cite{vynck_light_2023} whereas the onset of Anderson localization is commonly characterized by the Ioffe-Regel criterion $k_\mathrm{e}\ell_s \sim 1$.\cite{ioffe_noncrystalline_1960, wiersma_disordered_2013} 
Engineering the scattering mean free path $\ell_s$ of a medium is of practical importance  in fields such as lasing in random media,\cite{cao_lasing_2003} biomedical imaging,\cite{hohmann_direct_2021,goicoechea_reflection_2024} and remote sensing of turbid atmosphere and terrain.\cite{tsang_multiple_1980, picard_brief_2022}

The majority of previous formulas for $\ell_{s}$ have been derived to treat ordinary disordered particulate media consisting of identical dielectric particles embedded in a matrix of another dielectric constant. 
One such well-known formula is formulated as an integral that incorporates the Mie differential scattering cross-section $\dv*{\sigma_s}{\Omega}$ of a single spherical/circular particle,\cite{vandehulst_light_1957} but weighted with the structure factor $S(q)$ of all the scatterers (see definition in Sec. \ref{sec:def}):\cite{wolf_optical_1988, conley_light_2014, riboli_tailoring_2017,ma_light_2019, vynck_light_2023}
\begin{align} \label{eq:Mie-ells}
 \ell_s = \left[\rho \int_{\Omega_{d}} \dv{\sigma_s}{\Omega} (\theta;k,a,\varepsilon_2) S(2k_\mathrm{e} \sin(\theta/2)) \dd{\Omega} \right]^{-1},
\end{align}
where $\rho$ is the number density of scatterers, $\Omega_d$ stands for the entire solid angle in $d$-dimensional Euclidean space $\mathbb{R}^d$, and $\theta$ is the scattering angle.
Note that Eq. \eqref{eq:Mie-ells} is a particular form of a ``dependent-scattering" approximation, that is, one in which the scattering response explicitly depends on the spatial correlations between scatterers.
For brevity, we call Eq. \eqref{eq:Mie-ells} for $\ell_s$ the Mie estimate.\footnote{
While extensions of Eq.~\eqref{eq:Mie-ells} to polydisperse disk/sphere packings have been proposed,\cite{mishchenko_multiple_1992,vos_broadband_2013,yazhgur_light_2021} they are not considered here due to fundamental limitations. Independent-scattering approaches, which average single-particle cross-sections over the size distribution,\cite{mishchenko_multiple_1992,vos_broadband_2013} fail to recover the monodisperse limit of Eq.~\eqref{eq:Mie-ells}. Dependent-scattering formulations used in Ref. \cite{yazhgur_light_2021} are formally consistent with this limit but require explicit summation over all particles and cannot be expressed in terms of a microstructure-dependent descriptor (e.g., a structure factor) that provides clear physical insight.}

Importantly, the limited class of disordered particulate media for which Eq. \eqref{eq:Mie-ells} applies is only a very small subset of the infinite variety of possible two-phase microstructures.\cite{milton_theory_2002,torquato_random_2002}
%While some other formulas for $\ell_s$, based on \cite{vos_broadband_2013, mishchenko_light_1993} there have been several attempts to generalize Eq. \eqref{eq:Mie-ells} to polydisperse disk/sphere packings,  
For example, disordered two-phase media in $\R^d$ also include more general particulate media composed of arbitrarily shaped particles of various sizes as well as non-particulate media in which scattering cross-sections may be ill-defined, requiring another means to estimate the scattering mean free path.
Furthermore, such two-phase media can include ordinary disordered ones and those with correlated disorder.
Among the latter category, disordered hyperuniform two-phase composites \cite{zachary_hyperuniformity_2009,torquato_hyperuniform_2018,torquato_extraordinary_2022} are an emerging class of materials that are endowed with novel wave and other transport properties.\cite{lethien_enhanced_2017,xu_microstructure_2017, lopez_true_2018,yu_disordered_2018, torquato_multifunctional_2018, wang_hyperuniformity_2018, chen_designing_2018, lei_hydrodynamics_2019, gorsky_engineered_2019, nizam_dynamic_2021, klatt_wave_2022,kim_effective_2023,kim_theoretical_2024}

Disordered hyperuniform two-phase media are characterized by an anomalous suppression of volume-fraction fluctuations in the infinite-wavelength limit, that is, the spectral density $\fn{\tilde{\chi}_{_V}}{k}$ obeys the condition:\cite{zachary_hyperuniformity_2009,torquato_hyperuniform_2018}
\begin{align}
 \lim_{\abs{\vect{k}}\to 0} \fn{\tilde{\chi}_{_V}}{\vect{k}} = 0.
 \label{eq:HU-condition}
\end{align}
Hyperuniform media encompass all periodic systems, many perfect quasiperiodic media, and exotic disordered ones, generalizing our notions of long-range order to include exotic disordered varieties; see Ref. \cite{torquato_hyperuniform_2018} and references therein.
Disordered hyperuniform media lie between liquids and crystals; they are like liquids in that they are statistically isotropic without any Bragg peaks, and yet behave like crystals in the manner in which they suppress the large-scale density fluctuations. \cite{torquato_local_2003,zachary_hyperuniformity_2009,torquato_hyperuniform_2018}
Disordered {\it stealthy hyperuniform} (SHU) varieties are an important subclass defined as\cite{uche_constraints_2004, batten_classical_2008,zhang_ground_2015,torquato_ensemble_2015}
\begin{align} \label{eq:SHU}
 \tilde{\chi}_{_V}(\vect{k})=0 \qquad \text{for } 0<\abs{\vect{k}}<K,
\end{align} 
meaning that single scattering of incident radiation is completely suppressed for these wavevectors. \cite{batten_classical_2008,torquato_hyperuniform_2018}
These exotic disordered two-phase media exhibit novel electromagnetic wave transport properties, including complete photonic band-gap formation, high transparency in the optically dense regime, and maximized absorption.\cite{florescu_designer_2009, man_isotropic_2013, zhang_perfect_2016, leseur_highdensity_2016, zhang_transport_2016, ma_3d_2016, froufe-perez_band_2017, lopez_reciprocal_2017,gkantzounis_hyperuniform_2017,torquato_multifunctional_2018, chen_designing_2018,gorsky_engineered_2019, romero-garcia_stealth_2019, zhou_ultrabroadband_2020, sheremet_absorption_2020,sgrignuoli_hyperuniformity_2021,chen_hyperuniform_2021, christogeorgos_extraordinary_2021, yu_engineered_2021, granchi_nearfield_2022, scheffold_transport_2022, tavakoli_65_2022, cheron_wave_2022, klatt_wave_2022,kim_effective_2023,kim_extraordinary_2024, siedentop_stealthy_2024, kim_theoretical_2024,riganti_multiscale_2025}

In this work, we derive a new set of approximate formulas for the scattering mean free path $\ell_s$ of general two-phase dielectric media that accurately account for the microstructural information via the spectral density $\tilde{\chi}_{_V}(k)$ of the dielectric phases beyond the long-wavelength regime up to dimensionless wavenumber $k_1/s$ of the order of 1.
In contrast to the Mie estimate \eqref{eq:Mie-ells} that depends on the structure factor of particle centers, these formulas are applicable to more general particulate media and even non-particulate media, to which Eq. \eqref{eq:Mie-ells} cannot be directly applied.
Our more general approximations are extracted from the \textit{exact strong-contrast expansion} for the effective dynamic dielectric constant $\varepsilon_\mathrm{e}(k_1,\omega)$ of {\it non-absorbing} two-phase dielectric composites,\cite{torquato_nonlocal_2021,kim_effective_2023,kim_theoretical_2024}
expressed as 
\begin{align} \label{eq:ells-effective-medium}
 \ell_s = \qty{2 k_1 \mathrm{Im}[\sqrt{\varepsilon_\mathrm{e}(k_1,\omega)/\varepsilon_1} ]}^{-1},
\end{align}
where $\omega$ is the angular frequency, and $k_1$ is the incident wavenumber in a matrix phase with dielectric constant $\varepsilon_1$, as elaborated in Sec. \ref{sec:SCA}.
We present explicit formulas for layered media [i.e., effectively one-dimensional (1D)], transversely isotropic media [i.e., effectively two-dimensional (2D)] for both transverse electric (TE) and transverse magnetic (TM) polarizations, as well as three-dimensional (3D) fully statistically isotropic media for transverse polarization.
For such hyperuniform media, characterized by a power-law spectral density [i.e., $\tilde{\chi}_{_{V}}(q)\sim q^\alpha$ for small $q$], our formulas predict that the scattering mean free path exhibits a scaling behavior, that is, $\ell_{s} \sim {k_{1}}^{-(d+1+\alpha)}$, where $d$ is the space dimension.
In addition, our formulas predict that SHU media have infinitely large values of $\ell_s$ (or, equivalently, $1/\ell_s \approx 0$), indicating perfect transparency, for a finite-sized wavenumber range, that is, $0\leq k_1 \leq K_T$; see Eq. \eqref{eq:SHU-transparency}.

We also show that for the particulate media of identical spherical particles in two and three dimensions ($d=2,3$), these new predictions converge to the Mie estimate \eqref{eq:Mie-ells} in the weak-contrast regime (i.e., $\varepsilon_2/\varepsilon_1 \to 1^+$); see Sec. \ref{sec:long-wavelength}.
It is demonstrated by showing that for a circular or spherical particle, the quantity $\dv*{\sigma_s}{\Omega}$ becomes proportional to the {\it particle form factor}, denoted as $P(q;a)=\abs{\tilde{m}(q;a)}^2$, that is, the square of the Fourier transform of the particle indicator function.

Using new formulas based on Eq. \eqref{eq:ells-effective-medium} and the Mie estimate \eqref{eq:Mie-ells}, we estimate $\ell_s$ for five distinct models of two-phase media in both two and three dimensions (see Sec. \ref{sec:models}), characterized by a given volume fraction $\phi_{2}$ of the phase with a high dielectric constant $\varepsilon_2$.
These models include Debye random media,\cite{debye_scattering_1949,skolnick_understanding_2021} equilibrium hard disks/spheres,\cite{torquato_random_2002} hyperuniform $g_2$-invariant sphere packing,\cite{torquato_controlling_2002,wang_realizability_2022} hyperuniform polydisperse packing,\cite{kim_new_2019} and SHU packings.\cite{kim_ultradense_2025}
These models encompass both particulate and non-particulate types, spanning from typical non-hyperuniform media to hyperuniform and SHU media.
We utilize the inverse of {\it specific surface} (i.e., the mean interface area per volume) $s^{-1}$, as a natural characteristic inhomogeneity length scale to scale distance or wavenumbers, in order to compare properties of these models, as discussed in Ref. \cite{kim_characterizing_2021}.

We begin by verifying the accuracy of our predictions for $\ell_s$, demonstrating that these estimates for three selected models in two and three dimensions are in excellent agreement with the finite-difference time-domain (FDTD) simulations \eqref{eq:Mie-ells} for $k_{1}/s\lesssim 1$.
In this regime and for the special class of particulate media, the strong-contrast formulas are shown to be consistent with the predictions of the Mie estimates, but the former is more accurate than the latter for 2D TM polarization.
[Note that, for $d$-dimensional monodisperse sphere packings, the condition $k_{1}/s\lesssim 1$ translates to the diameter-to-wavelength ratio (i.e., $2a/\lambda=d\phi_2/\pi$) of around 0.5 or smaller.]
In what follows, we consider the relatively high dielectric contrast ratio $\varepsilon_{2}/\varepsilon_{1}=8$, which allows us to observe clearer discrepancies between the strong-contrast approximations and the Mie estimate beyond the long-wavelength regime. 
A dielectric contrast ratio of $8$ is achievable by real materials, as, for example, in rutile $\mathrm{TiO}{2}$/air systems for wavelengths of about 700~nm or to $\mathrm{Ge}/\mathrm{CaF}{2}$ systems for wavelengths near 5000~nm.\cite{polyanskiy_refractiveindexinfo_2024}
(Note that this contrast ratio is smaller than that of silicon, which has   a value of about 12.)
Our corresponding results for lower contrast ratios are even more accurate, since the strong-contrast
estimate generally improves in accuracy as $\varepsilon_{2}/\varepsilon_{1}$ approaches unity.\cite{torquato_nonlocal_2021}
We then compare the strong-contrast estimates for $\ell_s$ for all 2D and 3D models to investigate how $\ell_s$ varies with the microstructures.
The combination of our formulas with the methods to construct two-phase media with a targeted spectral density,\cite{chen_designing_2018,shi_computational_2023, shi_threedimensional_2025} lays the foundation for inversely designing novel wave characteristics in disordered hyperuniform and non-hyperuniform two-phase composites by engineering their spectral densities.
Such potential applications include transparent gradient index metamaterials,\cite{zhang_experimental_2019} random lasing media,\cite{gayathri_lasing_2023} and selective filtering materials that can be used for enhanced thermal insulation\cite{otanicar_filtering_2016} or structural colors.\cite{rothammer_tailored_2021}

In Sec. \ref{sec:def}, we define basic quantities.
We then derive our new formulas for $\ell_s$ in dimensions $d=2$ and $d=3$ and provide general remarks in Sec. \ref{sec:SCA}.
In Sec. \ref{sec:models}, we describe the model microstructures considered in this work.
Subsequently, we provide the results for these models in Sec. \ref{sec:results}.
Specifically, we numerically verify the accuracy of our predictions for $d=2,3$ in Sec. \ref{sec:FDTD} and \SI, respectively.
We then compare the predictions for all models in both dimensions $d=2$ and $d=3$ in Sec. \ref{sec:comp}.
Finally, we provide concluding remarks in Sec. \ref{sec:conclusions}.

\section{Definitions and Background}\label{sec:def}

\subsection{$n$-Point Correlation Functions and Spectral Density}\label{sec:two-phase}

A two-phase random medium is a domain of space of volume $V$ that is partitioned into two disjoint regions: a phase $i$ region with volume fraction $\phi_i$ for $i=1,2$.\cite{torquato_random_2002} 
The medium is fully statistically characterized by the $n$-point correlation $S^{(i)}_n({\bf x}_1,{\bf x}_2,\ldots,{\bf x}_n)$ 
associated with phase $i$ that gives the
probability of simultaneously finding $n$ points with positions ${\bf x}_1,{\bf x}_2,\ldots,{\bf x}_n$ in phase $i$.\cite{torquato_random_2002}

In this work, we are particularly interested in the one- and two-point correlation functions of \textit{statistically homogeneous} media, in which there is no preferred origin.
In such instances, the one-point correlation function is a constant, namely, $S_1^{(i)}(\vect{x})= \phi_i$, and the two-point correlation function depends on the relative displacement vector ${\bf r} \equiv {\bf x}_2-{\bf x}_1$ 
and hence $S_2^{(i)}({\bf x}_1,{\bf x}_2)=S_2^{(i)}({\bf r})$.\cite{torquato_random_2002,torquato_hyperuniformity_2016}
The autocovariance function $\chi_{_V}({\bf r})$ is defined as 
\begin{align}
 \chi_{_V}({\bf r}) = S^{(i)}_2({\bf r}) - {\phi_i}^2,
\end{align}
and thus is identical for each phase $i=1,2$.
At the extreme limits of its argument, $\chi_{_V}$ has the following asymptotic behavior
\begin{equation}
\chi_{_V}({\bf r}=0)=\phi_1\phi_2, \qquad \lim_{|{\bf r}| \rightarrow \infty} \chi_{_V}({\bf r})=0,
\label{eq:limits}
\end{equation}
the latter limit applying when the medium possesses no long-range order. 
If the medium is also statistically isotropic, then the autocovariance function ${\chi}_{_V}({\bf r})$ depends only on the magnitude of its argument $r=|\bf r|$,
and hence is a radial function. In such instances, its slope at the origin is directly related 
to the specific surface $s$; specifically, we have in any space dimension $d$, the asymptotic form, \cite{torquato_random_2002}
\begin{equation}
\chi_{_V}(r)= \phi_1\phi_2 - \beta(d) s \;r + {\cal O}(r^2),
\label{eq:auto-small-r}
\end{equation}
where $\beta(d)= \Gamma(d/2) /[2\sqrt{\pi} \Gamma((d+1)/2)].$

The nonnegative spectral density ${\tilde \chi}_{_V}({\bf k})$, which can be obtained from scattering experiments, \cite{debye_scattering_1949,debye_scattering_1957} is the Fourier transform of $\chi_{_V}({\bf r})$, that is,
\begin{equation}
{\tilde \chi}_{_V}({\bf k}) = \int_{\mathbb{R}^d} \chi_{_V}({\bf r}) e^{-i{\bf k \cdot r}} {\rm d} {\bf r} \ge 0, \qquad \mbox{for all} \; {\bf k}.
\end{equation}
For isotropic media, the spectral density only depends on $k=|{\bf k}|$ and is expressed as\cite{torquato_hyperuniform_2018}
\begin{align} \label{eq:spd}
 \tilde{\chi}_{_V}(k) & = (2\pi)^{d/2}\int_{0}^\infty r^{d-1} \chi_{_V}(r) \frac{J_{d/2 -1}(kr)}{(kr)^{d/2 -1}} \dd{r},
\end{align}
where $\fn{J_\nu}{x}$ is the Bessel function of the first kind of order $\nu$.
As a consequence of Eq. \eqref{eq:auto-small-r}, its decay in the large-$k$ limit is controlled
by the exact following power-law form with a coefficient proportional to the specific surface $s$\cite{torquato_disordered_2016}; specifically,
\begin{equation}
{\tilde \chi}_{_V}({\bf k}) \sim \frac{\gamma(d)\,s}{k^{d+1}}, \qquad k \rightarrow \infty,
\label{decay}
\end{equation}
where $\gamma(d)=2^d\,\pi^{(d-1)/2} \,\Gamma((d+1)/2)$ is a $d$-dimensional constant.

\subsection{Hyperuniformity and Local Volume Fraction Variance} \label{sec:hyper}

A hyperuniform two-phase medium defined by the condition \eqref{eq:HU-condition} is also characterized equivalently by a local volume-fraction variance $\fn{\sigma_V^2}{R}$ associated with a spherical window of radius $R$ that goes to zero asymptotically more rapidly than the inverse of the window volume (i.e., $R^{-d}$).\cite{zachary_hyperuniformity_2009,torquato_hyperuniform_2018}
Many disordered hyperuniform two-phase media can be characterized by a power-law scaling behavior of the spectral density in the small-wavenumber limit:
\begin{align}  \label{eq:spd-power-law}
 \tilde{\chi}_{_V}(\vect{k}) \sim \abs{\vect{k}}^\alpha  \qquad (\abs{\vect{k}} \to 0). 
\end{align}
Here, the value of a positive exponent $\alpha$ relates to the three different scaling regimes (classes) associated with the large-$R$ behaviors of $\sigma_V^2(R)$:\cite{zachary_hyperuniformity_2009,torquato_hyperuniform_2018}
\begin{align} 
\fn{\sigma^2_{_V}}{R}
\sim 
\begin{cases}
 R^{-(d+1)},   & \qquad \alpha >1 \qquad \text{(Class I)},\\
 R^{-(d+1)} \ln R, & \qquad \alpha = 1 \qquad \text{(Class II)},\\
 R^{-(d+\alpha)}, & 0 < \alpha < 1\qquad \text{(Class III)},
\end{cases}
\label{eq:HU-clases}
\end{align}
where $\alpha$ is a positive constant. 
Class I is the strongest form of hyperuniformity, which includes all perfect periodic media, SHU media (see Eq. \eqref{eq:SHU}), and other disordered ones.

By contrast, for any nonhyperuniform two-phase system, the variance has the following large-$R$ scaling behaviors:\cite{torquato_structural_2021}
\begin{align} \nonumber 
 \fn{\sigma^2_{_V}}{R}
\sim 
 \begin{cases}
 R^{-d}, &\quad~\quad~ \alpha = 0\quad \text{(typical nonhyperuniform)},\\
 R^{-d+\alpha},&-d<\alpha<0\quad\text{(antihyperuniform)},
 \end{cases} 
\end{align}
where $\alpha$ is defined in Eq. \eqref{eq:spd-power-law}.
A typical nonhyperuniform system has a positive and finite $\tilde{\chi}_{_V}(0)$, whereas an antihyperuniform one has an unbounded $\tilde{\chi}_{_V}(0)$ that is diametrically opposite to hyperuniform systems.
Antihyperuniform systems include systems at thermal critical points (e.g., liquid-vapor and magnetic critical points),\cite{stanley_introduction_1987, binney_theory_1992} fractals,\cite{mandelbrot_fractal_1982} disordered nonfractals,\cite{torquato_local_2022} and certain substitution tilings. \cite{oguz_hyperuniformity_2019}

\subsection{Particulate Medium of Identical Particles} \label{sec:Sk}

For the special case of particulate media (or packing\cite{torquato_perspective_2018}) of identical `nonoverlapping' spheres of radius $a$ that comprises phase 2 with packing fraction $\phi_2 = \rho v_1(a)$, the spectral density is simply related to the structure factor:\cite{torquato_microstructure_1985,torquato_random_2002}
\begin{align}
{\tilde \chi}_{_V}({\bf k})&= \rho P(k;a) S({\bf k}).
\label{eq:chi_V-S}
\end{align}
where $v_1(a)$ is the volume of a $d$-dimensional sphere of radius $a$, $P(k;a)$ is the particle form factor\cite{vandehulst_light_1957} of a $d$-dimensional sphere of radius $a$, expressed as
\begin{equation} \label{eq:p-formfactor}
P(k;a)= \left(\frac{2\pi a}{k}\right)^{d} J_{d/2}^2(ka).
\end{equation}
Since $P(k;a)$ is a positive, bounded, well-behaved function around $k=0$, it immediately follows from expression (\ref{eq:chi_V-S}) that if the underlying
point process is hyperuniform with a power-law structure factor $S({\bf k}) \sim |{\bf k}|^{\alpha}$ in the limit $|{\bf k}|\to 0$, then
the spectral density ${\tilde \chi}_{_V}({\bf k})$ inherits the same power-law form only through the structure factor, not $P(k;a)$.\cite{torquato_hyperuniformity_2016}
Moreover, it is clear that for SHU packings, relation \eqref{eq:chi_V-S} dictates that both ${\tilde \chi}_{_V}({\bf k})$ and $S({\bf k})$ are zero for the same wavenumbers.
However, the spectral density also vanishes at the zeros of the function $P(k;a)$.\cite{torquato_hyperuniformity_2016} %, which is determined by the zeros of $J_{d/2}(ka)$

\subsection{Differential Scattering Cross-Section in Mie Theory}\label{sec:Mie}

Here, we provide well-known formulas for the Mie differential scattering cross-section\cite{vandehulst_light_1957} $\dv*{\sigma_s}{\Omega}$ for a cylinder and sphere used in Eq. \eqref{eq:Mie-ells}.
We begin with the case of an infinitely long dielectric cylinder of radius $a$ and dielectric constant $\varepsilon_2$ embedded in a matrix of dielectric constant $\varepsilon_1$.
In this case, the expression for $\dv*{\sigma_s}{\Omega}$ for the TE and TM polarizations is given as\cite{vandehulst_light_1957}
\begin{align}
 \dv{\sigma_{s}}{\theta} =& \frac{2}{\pi k_{1}} \abs{T(\theta)}^2 \nonumber \\
 =& 
 \frac{2}{\pi k_{1}} \times
 \begin{cases}
 \abs{a_{0} + 2\sum_{n=1}^\infty a_{n} \cos(n\theta)}^2, & \text{TE}\\
 \abs{b_{0} + 2\sum_{n=1}^\infty b_{n} \cos(n\theta)}^2, & \text{TM},
 \end{cases} \label{eq:diff-sigma-cylinder} 
\end{align}
where the coefficients are given as
\begin{align*} %\label{eq:an-cylinder}
 a_{n}\equiv& \frac{J_{n}'(y)J_{n}(x)-m J_{n}(y)J_{n}'(x)}{J_{n}'(y)H_{n}^{(2)}(x)-mJ_{n}(y){H_{n}^{(2)}}'(x)}, \\
 b_{n}\equiv& \frac{mJ_{n}'(y)J_{n}(x)- J_{n}(y)J_{n}'(x)}{mJ_{n}'(y)H_{n}^{(2)}(x)-J_{n}(y){H_{n}^{(2)}}'(x)}, %\label{eq:bn-cylinder}
\end{align*}
where $H_{n}^{(2)}(x)$ is the Hankel function of the second kind of order $n$, $m\equiv\sqrt{\varepsilon_2/\varepsilon_1}$ is the refractive index, $x\equiv k_1 a$ and $y\equiv m x$ are the dimensionless wavenumbers in the matrix and the cylinder, respectively.

The analogous expression for a sphere of radius $a$ is expressed as\cite{vandehulst_light_1957}
\begin{align}
\dv{\sigma_{s}}{\Omega} =& \frac{1}{2{k_{1}}^2} [\abs{T_{1}(\theta)}^2+\abs{T_{2}(\theta)}^2] 
\nonumber \\
=& 
\frac{1}{2{k_{1}}^2} \times
\Bigg[\abs{\sum_{n=1}^\infty \frac{2n+1}{n(n+1)} [a_{n}\fn{\pi_{n}}{\cos \theta} + b_{n}\fn{\tau_{n}}{\cos \theta}] }^2 \nonumber \\
&+ \abs{\sum_{n=1}^\infty \frac{2n+1}{n(n+1)} [b_{n}\fn{\pi_{n}}{\cos \theta} + a_{n}\fn{\tau_{n}}{\cos \theta}] }^2 \Bigg]
\label{eq:diff-sigma-sphere} ,
\end{align}
where the angle-dependent functions $\pi_n$ and $\tau_n$ are expressed in terms of the associated Legendre polynomials $P_l^m (x)$: 
\begin{align*}
\fn{\pi_{n}}{\cos \theta}\equiv & \frac{1}{\sin \theta} \fn{P_{n}^1}{\cos \theta } %\label{eq:pi-n} 
\\
\fn{\tau_{n}}{\cos \theta}\equiv & \dv{\theta} \fn{P_{n}^1}{\cos \theta}, %\label{eq:tau-n}
\end{align*}
and 
\begin{align*}
a_{n} =& \frac{\psi_{n}'(y)\psi_{n}(x)-m \psi_{n}(y)\psi_{n}'(x)}{\psi_{n}'(y)\zeta_{n}(x)-m \psi_{n}(y)\zeta_{n}'(x)}, \\
b_{n} =& \frac{m\psi_{n}'(y)\psi_{n}(x)-\psi_{n}(y)\psi_{n}'(x)}{m\psi_{n}'(y)\zeta_{n}(x)- \psi_{n}(y)\zeta_{n}'(x)}, 
\end{align*}
where $\psi_{n}(x)\equiv x j_{n}(x)$ and $\zeta_{n}(x)\equiv x h_{n}^{(2)}(x)$, and $j_{n}(x)$ and $h_{n}^{(2)}(x)$ are the spherical Bessel function and spherical Hankel function, respectively.
Here, the three parameters $m$, $x$, $y$ are the same as those used for the cylinder.

\section{Formulas from Strong-Contrast Approximations}
\label{sec:SCA}

We consider a macroscopically large ellipsoidal specimen / sample of a medium that is statistically homogeneous and consists of two non-absorbing phases, embedded within an infinitely large reference phase with a dielectric constant tensor $\varepsilon_1$.\cite{torquato_nonlocal_2021}
The shape of this specimen is purposely chosen to be nonspherical, since any rigorously correct expression for the effective property must ultimately be independent of the shape of the macroscopic composite specimen in the infinite-volume limit,\footnote{Interestingly, it has been numerically verified that the shape-independence of the effective property is still valid for a large finite-sized specimen.\cite{gower_effective_2021, gower_model_2023}} which was shown to be the case in Ref. \cite{torquato_nonlocal_2021}.
The medium is subjected to an applied electric wave of an incident wavevector $\vb{k}_1$ and a given polarization vector.
Under these assumptions, the exact strong-contrast series for \textit{nonlocal} effective dynamic dielectric constants $\varepsilon_\mathrm{e}(\vb{k}_q)$ are obtained as\cite{torquato_nonlocal_2021, kim_effective_2023, kim_theoretical_2024}
\begin{align}
&\phi_p \tens{L}_p^{(q)}\cdot 
 \left(
 \left\{\tens{I} + \tens{D}^{(q)} \cdot \left[\fn{\tens{\varepsilon}_\mathrm{e}}{\vect{k}_q} - {\varepsilon}_q \tens{I} \right] \right\} \cdot \left[\fn{\tens{\varepsilon}_\mathrm{e}}{\vect{k}_q} - {\varepsilon}_q \tens{I} \right]^{-1} 
 \right) 
 \cdot \phi_p \tens{L}_p^{(q)}
 \nonumber \\
& \qquad =
 \phi_p \tens{L}_p^{(q)} 
 - \sum_{n=2}^\infty \fn{\tens{\mathcal{A}}_n^{(p)}}{\vect{k}_1; S_1^{(p)},...,S_n^{(p)}} ,
\label{eq:str-exp-trunc}
\end{align}
where $q(=1)$ indicates the matrix phase, $p(=2)$ indicates the polarized phase, $\fn{\tens{\mathcal{A}}_n^{(2)}}{\vect{k}_1; S_1^{(2)},...,S_n^{(2)}}$ is a wavevector-dependent second-rank tensor that is a functional involving the set of correlation functions $S_1^{(2)}, S_2^{(2)}, \ldots, S_n^{(2)}$, defined in Sec.~\ref{sec:two-phase}, and products of the principal part of the dyadic Green's function.
The linear fractional form of the series \eqref{eq:str-exp-trunc} makes it converge rapidly, and thus its lower-order truncations lead to accurate approximate formulas for $\fn{\tens{\varepsilon}_e}{\vect{k}_q, \omega}$, even for large contrast ratios. 
This is to be contrasted with standard weak-contrast expansions that do not converge rapidly for large contrast ratios; see \cite{kim_theoretical_2024} for a quantitative explanation.
The tensor $\tens{D}^{(1)}$ is related to an infinitesimal region around the singularity in the Green's function, \cite{torquato_nonlocal_2021, yaghjian_electric_1980} taken according to the structural symmetry of a medium.\cite{kim_effective_2023, kim_theoretical_2024}
Importantly, the effective tensor in Eq. \eqref{eq:str-exp-trunc} accounts for nonlocal effects\cite{agranovich_crystal_1984} such that the corresponding homogenized constitutive relation is nonlocal in space, resulting in a series expansion that is valid and accurate beyond the long-wavelength regime.\cite{torquato_nonlocal_2021}
The reader is referred to Refs. \cite{torquato_nonlocal_2021, kim_effective_2023, kim_theoretical_2024} for derivations.

It is instructive to clarify the physical meaning of the strong-contrast expansion \eqref{eq:str-exp-trunc}.
This approach describes the effective dielectric response of an infinitesimal volume element of prescribed shape, represented by the tensor $\tens{D}^{(1)}$, located at position $\vect{x}$ to the coherent field at $\vect{x}$ generated by the surrounding composite medium, which excludes the self-field of the volume element.
In the full series given in Eq. \eqref{eq:str-exp-trunc}, this coherent field systematically incorporates near-field effects arising from neighboring particles, as well as the influence of long-range density fluctuations encoded in the microstructure.
At the two-point level considered here, the strong-contrast approximation partially accounts for such near-field effects through contributions from the surrounding volume elements of the dielectric phase located at positions $\vect{x}' \neq \vect{x}$, whose spatial separations $\vect{x}-\vect{x}'$ are characterized by the two-point correlation function $S_2^{(2)}(\vect{x}-\vect{x}')$, or equivalently by the associated spectral density $\tilde{\chi}_{_V}(\vect{k})$, while correlations among these surrounding volume elements themselves are neglected.
As a result, these approximations are applicable to general two-phase media and capture long-range spatial correlations, while only implicitly resolving particle-specific modes in the short-wavelength regime.
Indeed, truncations of these expansions at the two-point level are resummations of the strong-contrast expansions that still accurately
capture multiple scattering to all orders via the microstructural information embodied in the spectral
density.\cite{torquato_nonlocal_2021}

The estimate of $\ell_s$ obtained from Eq. \eqref{eq:str-exp-trunc} therefore differs fundamentally from the Mie estimate \eqref{eq:Mie-ells}. 
The latter is built on the exact single-particle scattering response, expressed through the differential scattering cross-section $\pdv*{\sigma_s}{\Omega}$, and treats each particle as being excited by the same incident plane wave.
Within this framework, positional correlations enter only through far-field interference effects in the scattered intensity.\footnote{In the radiative-transfer limit, the Mie estimate \eqref{eq:Mie-ells} can also be obtained from the angular integral of the intensity vertex in the Bethe-Salpeter equation.\cite{vynck_light_2023}} 
Consequently, while the Mie estimate accurately captures particle-specific resonances in the short-wavelength regime, it neglects near-field or recurrent-scattering effects associated with strong local-field coupling between neighboring particles, which can lead to substantial errors in dense or strongly correlated media.\cite{rezvaninaraghi_nearfield_2015,pattelli_role_2018,ma_light_2019}

By selecting an appropriate form for $\tens{D}^{(1)}$ and truncating the series at the two-point level, one obtains approximate formulas involving $\tilde{\chi}_{_V}(k)$ that show great accuracy.
In what follows, we state these key formulas for the effective dielectric constants for (i) layered, (ii) transversely isotropic, and (iii) fully statistically isotropic media to derive corresponding formulas for the scattering mean free path $\ell_s$ given in Eq. \eqref{eq:ells-effective-medium}. 
We apply these formulas for $d=2,3$ to estimate $\ell_s$ of individual models in Sec. \ref{sec:results}.

\subsection{Layered Media} \label{sec:layered}

A layered two-phase medium in three dimensions consists of infinite parallel slabs of phases 1 and 2 whose thicknesses are derived from a prescribed 1D two-phase model.
We consider the situation in which vector waves are incident parallel to the symmetry axis $\uvect{z}$ of such a layered medium, that is, $\vect{k}_1 = k_1 \uvect{z}$.
Due to its symmetries, one can decompose $\fn{\tens{\varepsilon}_\mathrm{e}}{k_1}$ into two orthogonal components: $\fn{\tens{\varepsilon}_\mathrm{e}}{k_1} = \fn{\varepsilon_\mathrm{e} ^\perp}{k_1} \left(\tens{I}-\uvect{z}\uvect{z} \right) + \fn{\varepsilon_\mathrm{e}^z}{k_1}\uvect{z}\uvect{z}$, where $\fn{\varepsilon_\mathrm{e} ^\perp}{k_1}$ and $\fn{\varepsilon_\mathrm{e} ^z}{k_1}$ are for the transverse and longitudinal polarizations, respectively.
(The reader is referred to Ref. \cite{kim_effective_2023} for derivations.)
Since propagating waves occur only for the transverse polarization here, we focus on the {\it scaled} strong-contrast approximation at the two-point level for $\fn{\varepsilon_\mathrm{e} ^\perp}{k_1}$, given by 
\begin{align}
    \frac{\fn{\varepsilon_\mathrm{e}^\perp}{k_1}}{\varepsilon_1} =& 
        1 + \frac{{\phi_2}^2 \varepsilon_2/\varepsilon_1\BETA{1}{21}}{\phi_2 - (\varepsilon_2 \BETA{1}{21}) \fn{A_2^\perp}{k_*; \expval{\varepsilon}} },
    \label{eq:layered}
\end{align}
where $k_*\equiv k_1 \sqrt{\expval{\varepsilon}/\varepsilon_1}$, $\BETA{1}{pq}\equiv  1-\varepsilon_q/\varepsilon_p$ is the 1D counterpart of the \emph{dielectric polarizability}, and the second-order term is expressed as $\fn{A_2^\perp}{k; \varepsilon}\equiv \F{1}{k}/\varepsilon$.
Here, $\F{1}{k}$ is the {\it nonlocal attenuation function} for 1D two-phase media,\cite{kim_effective_2023} defined as
\begin{align}
    \F{1}{k}
    \equiv & 
    \frac{{k}^2}{\pi} \mathrm{p.v.}\int_{0}^{\infty} \dd{q} 
    \frac{\tilde{\chi}_{_V}(q)}{{q}^2 - {(2k)}^2}
    \nonumber \\
    &+
    \frac{i k}{4} [
    \tilde{\chi}_{_V}(0) +  \tilde{\chi}_{_V}(2k)
    ]   \label{eq:F-1D},    
\end{align}
where $\mathrm{p.v.}$ stands for the Cauchy principal value. 
Note that function $\F{d}{k}$ is a dimensionless and complex-valued quantity that depends solely on the microstructure for any space dimension $d$ via the spectral density.
In the static case (i.e., $k_1=0$), Eq. \eqref{eq:layered} reduces to the arithmetic mean of the local dielectric constants,
$
\fn{\varepsilon_\mathrm{e}^\perp}{0} = \expval{\varepsilon}\equiv \phi_1 \varepsilon_1 + \phi_2 \varepsilon_2 ,
$
which is exact for any 1D microstructure.\cite{torquato_random_2002}

\subsection{Transversely Isotropic Media} \label{sec:2D-theory}

A transversely isotropic medium in three dimensions consists of infinite parallel dielectric cylinders of phase 2 embedded in phase 1 whose cross-section is derived from 2D statistical isotropic media.
We consider the situation in which vector waves are normally incident on the symmetry axis $\uvect{z}$ of transversely isotropic media. 
Due to the symmetries of the problems, one can decompose the effective dielectric constant tensor into two orthogonal components $\ETM{k_1}$ and $\ETE{k_1}$ for TM and TE polarizations, respectively, as follows:
$\fn{\tens{\varepsilon}_\mathrm{e}}{k_1} = \ETM{k_1} \uvect{z}\uvect{z} + \ETE{k_1}(\tens{I}-\uvect{z}\uvect{z})$.
Detailed derivation is provided in Ref. \cite{kim_theoretical_2024}.
The corresponding \emph{scaled} strong-contrast approximations at the two-point level are
\begin{align}
 \frac{\ETM{k_1}}{\varepsilon_1} 
 =& 
 1+ \frac{ 
 {\phi_2}^2 \qty[(\varepsilon_2+\varepsilon_1) \BETA{2}{21}]
 } 
 {\phi_2 - \ATM{2}{k_*^{TM}, \expval{\varepsilon}} \qty[(\varepsilon_2+\varepsilon_1) \BETA{2}{21}]
 }, \label{eq:trans-TM}
 \\
 \frac{\ETE{k_1}}{\varepsilon_1}
 =& 
 1+ \frac{ 
  2 {\phi_2}^2 \BETA{2}{21}
 }
 { 
 \phi_2(1-\phi_2 \BETA{2}{21}) 
 - 
 \ATE{2}{k_*^{TE}; \varepsilon_{BG}^{(2D)}} \qty[2\varepsilon_1 \BETA{2}{21}]
 }
  \label{eq:trans-TE} , 
\end{align}
where $\BETA{2}{pq}\equiv (\varepsilon_p - \varepsilon_q)/(\varepsilon_p + \varepsilon_q)$ is the 2D counterpart of the dielectric polarizability, and $k_*^{TE} \equiv k_1 \sqrt{\varepsilon_{BG}^{(2D)}/\varepsilon_1}$ and $k_*^{TM} \equiv k_1 \sqrt{\expval{\varepsilon}/\varepsilon_1}$ are wavenumbers in the optimal reference phase for TE and TM polarizations, respectively, and $\varepsilon_{BG}^{(2D)}$ is the Bruggeman approximation for 2D two-phase media. \cite{bruggeman_berechnung_1935,torquato_random_2002}
Here, the second-order coefficients for different polarizations are expressed as $\fn{A_2^{TM}}{k; \varepsilon}=2 \fn{A_2^{TE}}{k; \varepsilon}=- \pi/(2 \varepsilon) \F{2}{k}$, where $\F{2}{k}$ is the nonlocal attenuation function for 2D statistically isotropic two-phase media,\cite{torquato_nonlocal_2021} defined as
\begin{align}
 \F{2}{k}
 =&
 \frac{-1}{\pi}
 \Bigg\{
 \frac{{k}^2}{\pi^2}
 \int_{0}^{\pi/2} \dd{\phi} 
 \qty[ \mathrm{p.v.} 
 \int_0^\infty \dd{q} \frac{2q\tilde{\chi}_{_V}(q)}{q^2-(2k\cos\phi)^2}
 ]
 \nonumber \\
 &+ i\frac{{k}^2}{\pi} \int_0^{\pi/2} \tilde{\chi}_{_V}(2k \cos\phi)\dd{\phi}
 \Bigg\}.
 \label{eq:F-trans-Fourier}
\end{align}
In the static limit (i.e., $k_1\to0$), Eqs. \eqref{eq:trans-TM} and \eqref{eq:trans-TE} reduce to the arithmetic means of the local dielectric constant and the Maxwell-Garnett approximation for $d=2$,\cite{torquato_random_2002} respectively:
 \begin{align}
 \ETM{0} =& \expval{\varepsilon} =\varepsilon_1\phi_1 + \varepsilon_2 \phi_2, %\label{eq:trans-TM-static}
 \nonumber \\
 \ETE{0} 
 =& 
 \varepsilon_q
 \frac{
 1+\phi_p\BETA{2}{pq}}
 {
 1-\phi_p\BETA{2}{pq}
 }.
 \nonumber  
  %\label{eq:trans-TE-static} 
 \end{align}

\subsection{Fully Statistically Isotropic Media}\label{sec:3D-theory}

For fully statistically isotropic media, which we call 3D (isotropic) media for brevity, the effective dielectric constant is independent of the direction of the incident vector wave.
Additionally, it is sufficient to consider the transverse polarization alone, as the longitudinal contribution is negligible.
Then, the strong-contrast approximation at the two-point level (see Ref. \cite{torquato_nonlocal_2021} for derivation) is given as
\begin{align}
\frac{\fn{\varepsilon_\mathrm{e}}{k_1}}{\varepsilon_1} &= 1+ \frac{3 {\phi_2} ^2 \BETA{3}{21} }{\phi_2(1-\phi_2 \BETA{3}{21}) - \fn{A_2^{(2)}}{k_*^{\mathrm{(3D)}};\varepsilon_\mathrm{BG}^{\mathrm{(3D)}}} [3\varepsilon_1\BETA{3}{21}]},
\label{eq:iso-2pt}
\end{align}
where $\BETA{3}{pq}\equiv (\varepsilon_p - \varepsilon_q)/(\varepsilon_p + 2\varepsilon_q)$ is the 3D dielectric polarizability, $\varepsilon_{BG}^{(3D)}$ is the Bruggeman approximation for 3D two-phase media,\cite{bruggeman_berechnung_1935,torquato_random_2002} and $k_*^\mathrm{(3D)} \equiv k_1 \sqrt{\varepsilon_{BG}^{(3D)}/\varepsilon_1}$.
Here, the second-order coefficient is $A_2^{(2)}(k;\varepsilon) = -\sqrt{2\pi}/(3\varepsilon) \F{3}{k}$, where $\F{3}{k}$ is the nonlocal attenuation function for 3D statistically isotropic two-phase media,\cite{torquato_nonlocal_2021} defined as
\begin{align}
 \F{3}{k}
 =&
 -\frac{4k^2}{(2\pi)^{5/2}} \Bigg\{\int_{0}^1 \dd{x} \qty[\mathrm{p.v.}\int_{0}^\infty \dd{q} \frac{q^2\tilde{\chi}_{_V}(q)}{q^2-(2xk)^2}] 
 \nonumber \\
&+ \frac{\pi}{4k} \int_0 ^{2k}\dd{q'} q'\tilde{\chi}_{_V}(q') \Bigg\}.
\label{eq:F-iso-Fourier}
\end{align}
In the static limit (i.e., $k_1\to0$), Eq. \eqref{eq:iso-2pt} reduces to the Maxwell-Garnett approximation\cite{torquato_random_2002} for $d=3$:
 \begin{align}
 \varepsilon_\mathrm{e}(0) 
 =& 
 \varepsilon_q
 \frac{
 1+2\phi_p\BETA{3}{pq}}
 {
 1-\phi_p\BETA{3}{pq}
 }.
 \nonumber  
 \end{align}

\subsection{Perfect Transparency Interval in SHU Media}
\label{sec:transparency}

The strong-contrast approximations given in Eqs. \eqref{eq:layered}, \eqref{eq:trans-TM}, \eqref{eq:trans-TE}, and \eqref{eq:iso-2pt} predicted that a disordered SHU medium [defined by Eq. \eqref{eq:SHU}] has a perfect transparency interval,\cite{kim_effective_2023,kim_theoretical_2024,kim_extraordinary_2024} that is, zero imaginary part of the effective dielectric constant for the following range of wavenumbers:
\begin{align} \label{eq:SHU-transparency}
 0\leq k_1 \leq K_T \equiv \frac{K}{2 \sqrt{\varepsilon_*/\varepsilon_1}},
\end{align}
where 
\begin{align*}
\varepsilon_* =
\begin{cases}
 \expval{\varepsilon}, & \text{1D or 2D TM},\\
 \varepsilon_\mathrm{BG}^\mathrm{(2D)}, & \text{2D TE},\\
 \varepsilon_\mathrm{BG}^\mathrm{(3D)}, & \text{3D}.\\
\end{cases} 
\end{align*}
Such perfect transparency is easily proven by showing that the second-order coefficient $A_2$ [or, equivalently, $\F{d}{k}$] is real-valued within this range.
It has also been analytically shown that the interval remains valid up to the three-point level for layered media (1D) and TM polarization in transversely isotropic media (2D).\cite{kim_theoretical_2024, kim_extraordinary_2024} 
These results imply that $\ell_s$ for a SHU medium is infinitely large or, at least, sufficiently larger than a sample of practical size in the interval \eqref{eq:SHU-transparency}.
It is worth noting that neglecting nonlocal effects in strong-contrast approximations can lead to an overestimation of $K_T$ by approximately a factor of two, as discussed in the Supplementary Material of Ref. \cite{torquato_nonlocal_2021}.

\subsection{Long-Wavelength Regime}
\label{sec:long-wavelength}

Here, we consider the predictions of $\ell_s$ in the long-wavelength regime, that is, $k_1/s \ll 1$.
In this regime, we begin with the imaginary part of the effective dielectric constant that is determined by the asymptotic behavior of the imaginary part of the second-order coefficients $A_2$ or, equivalently, the attenuation function, regardless of the structural symmetry and polarizations.\cite{torquato_nonlocal_2021,kim_theoretical_2024,kim_extraordinary_2024}
Rewriting the imaginary part of $F^{\mathrm{(dD)}}(k)$ as an angular integral similar to the Mie estimate \eqref{eq:Mie-ells}, we obtain
\begin{align} \label{eq:im-eps}
 \mathrm{Im}[\varepsilon_\mathrm{e}(k_1)] 
&\approx 
 b' {k_1}^d\int_{\Omega_d} \tilde{\chi}_{_V}(2k_* \sin(\theta/2))\dd{\Omega} ,
\end{align}
where $k_*$ stands for the wavenumber in the optimal reference phase explained in Secs. \ref{sec:layered}, \ref{sec:2D-theory}, and \ref{sec:3D-theory}, and $b'$ is a constant coefficient depending on both space dimension $d$ and polarization, given by 
\begin{align}
 b' =
 \begin{cases}
 \frac{1}{4\sqrt{\expval{\varepsilon}}} [\varepsilon_2 \BETA{1}{21}]^2,&    d=1,\\
 \frac{1}{8\pi}
 \left[ \qty(1+\frac{\varepsilon_{2}}{\varepsilon_{1}})\BETA{2}{21} \right]^2, & d=2, \text{ TM}
 \\
 \frac{1}{16\pi} \qty[\frac{2\BETA{2}{21}}{1-\phi_2\BETA{2}{21}}]^2, & d=2, \text{ TE}
 \\
 \frac{1}{24\pi^2}
 \left[\frac{3 \BETA{3}{21}}{1-\phi_2 \BETA{3}{21}}\right]^2 \sqrt{ \varepsilon_{\mathrm{BG}}^{\mathrm{(3D)}} }, & d=3
 \end{cases}. \label{eq:coeff}
\end{align}
Substituting Eq. \eqref{eq:im-eps} into Eq. \eqref{eq:ells-effective-medium} gives a simplified expression:
\begin{align}
&\ell_s (k_1) \approx 
 \left[\frac{k_1 \mathrm{Im}[\varepsilon_\mathrm{e}(k_1)]}{\sqrt{\varepsilon_\mathrm{e}(0)/\varepsilon_1} } \right]^{-1} \nonumber 
\\
 \approx&
 \left[\int_{\Omega_d} \frac{b' {k_1}^{d+1}}{\sqrt{\varepsilon_\mathrm{e}(0)/\varepsilon_1}} \tilde{\chi}_{_V}(2k_* \sin(\theta/2))\dd{\Omega}\right]^{-1}.
\label{eq:ells_SCA}
\end{align}
Note that in this regime where the scattering is also weak (i.e., $k_* \ell_s \gg 1$), the imaginary part of the transverse component of the {\it self-energy},\cite{vynck_light_2023} denoted by $\Im[\Sigma_\perp]$, in a random medium can be approximated using Eq. \eqref{eq:ells_SCA} as follows:
\begin{align} \label{eq:self-energy}
 \mathrm{Im}[\Sigma_\perp(k_*)] 
 \approx & 
 k_* /\ell_s 
 \nonumber \\
 \approx &
 \int_{\Omega_d} b' {k_1}^{d+2} \tilde{\chi}_{_V}(2k_* \sin(\theta/2))\dd{\Omega}, 
\end{align}
the right-handed side of which is proportional to the term $-{k_1}^2\mathrm{Im}[\F{d}{k_1}]$.
Thus, $\F{d}{k}$ is as important as $\Sigma$ to understand the effective wave behaviors in complex random media.
For two-phase media, characterized by a power-law spectral density [i.e., $\tilde{\chi}_{_{V}}(k)\sim k^\alpha$ for small $k$], Eq. \eqref{eq:im-eps} directly gives the following scaling behavior:\cite{torquato_nonlocal_2021,kim_theoretical_2024,kim_extraordinary_2024}
\begin{align}
 \mathrm{Im}[\varepsilon_\mathrm{e}(k_1)] 
 \sim 
 k_1 ^{d+\alpha}. \nonumber
\end{align}
Substituting this expression into Eq. \eqref{eq:ells_SCA} yields
\begin{align} 
 \ell_s (k_1) 
 \sim & k_1^{-(d+1+\alpha)}.\label{eq:ells_scaling}
\end{align}
Here, we note that 3D nonhyperuniform media (i.e., $\alpha=0$) exhibit $1/\ell_s = \mu_s \sim k_1^{4}$, which is consistent with the exponent associated with Rayleigh scattering.\cite{andreev_scattering_1978, shepilov_anomalous_2023}

In the rest of this subsection, we focus on particulate media consisting of identical spherical particles of radius $a$ throughout a matrix in order to confirm the consistency between our predictions and the Mie estimate \eqref{eq:Mie-ells} for $d=2,3$.
The Mie estimate \eqref{eq:Mie-ells} also yields the scaling \eqref{eq:ells_scaling} in the long-wavelength regime with a coefficient that is equal to those of the strong-contrast estimates in the weak-contrast regime; however, this coefficient deviates from those of the strong-contrast estimates as the contrast ratio increases.
This behavior can be easily shown by approximating $\dv*{\sigma_s}{\Omega}$ in Eq. \eqref{eq:Mie-ells} with the Rayleigh scattering cross-section\cite{vandehulst_light_1957,jackson_classical_1999} that is proportional to ${k_1}^{d+1}$ for $d=2,3$.
Importantly, by substituting Eq. \eqref{eq:chi_V-S} to Eq. \eqref{eq:ells_SCA}, one obtains
\begin{align} \label{eq:ells_SCA_formfactor}
 \ell_s \approx \left[ \rho\int_{\Omega_d} 
 f(k_1,\theta; \varepsilon_2/\varepsilon_1) S(2k_* \sin(\theta/2))\dd{\Omega}\right]^{-1},
\end{align}
where we call $f(k_1,\theta; \varepsilon_2/\varepsilon_1)$ a differential {\it weighted} particle form factor of a $d$-dimensional dielectric sphere, expressed as 
\begin{align} \label{eq:form-vs-sigma}
 f(k_1,\theta; \varepsilon_2/\varepsilon_1) \equiv 
 \frac{b' {k_1}^{d+1}}{\sqrt{\varepsilon_\mathrm{e}(0)/\varepsilon_1}} P(2k_* \sin(\theta/2); a) ,
\end{align}
where the particle form factor $P(q;a)$ is given in Eq. \eqref{eq:p-formfactor}.
Comparing Eq. \eqref{eq:ells_SCA_formfactor} to the Mie estimate \eqref{eq:Mie-ells}, one can observe that the quantity $f(k_1,\theta; \varepsilon_2/\varepsilon_1)$ plays the same role as the differential scattering cross-section $\dv*{\sigma_s}{\Omega}$.
Indeed, the term \eqref{eq:form-vs-sigma} corresponds to the leading-order contribution of $\dv*{\sigma_s}{\Omega}$ in the weak-contrast regime $\varepsilon_2/\varepsilon_1 \to 1^+$; see Appendix \ref{app:FormFactor-Mie} for details.
Specifically, the leading-order term in $\BETA{d}{21}$ of the Mie total scattering cross-section $\sigma_s$ is identical to the angular integral of Eq. \eqref{eq:form-vs-sigma} for 2D TE polarization and 3D transverse one.
For 2D TM polarization, the coincidence is more dramatic: the leading-order term in $(\varepsilon_2-\varepsilon_1)$ of $\dv*{\sigma_s}{\Omega}$ from the Mie theory is identical to Eq. \eqref{eq:form-vs-sigma} for all $\theta$ and $k_1a$.

\begin{figure*}
\includegraphics[width=0.95\textwidth]{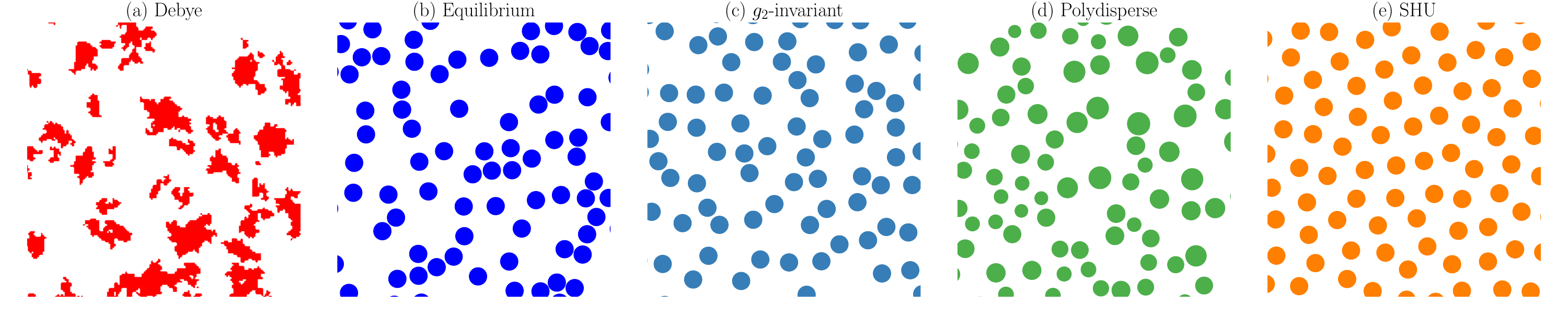}
\caption{Representative images of five models of 2D two-phase media with $\phi_2=0.25$: Debye random media (a), equilibrium packing of hard disks (b), hyperuniform $g_2$-invariant packing (c), hyperuniform polydisperse packing (d), and SHU packing (e).
In each panel, the colored regions depict the polarized phase (phase 2).
All images are scaled to have $1/s=1$.
Panel (a) is reproduced from data in Ref. \cite{skolnick_understanding_2021}.
\label{fig:model}}
\end{figure*}

\section{Model Microstructures}\label{sec:models}

We study five models of 2D and 3D disordered two-phase microstructures of transversely isotropic and fully isotropic materials.
These models include non-hyperuniform media, nonstealthy hyperuniform ones, and SHU ones.
Figure \ref{fig:model} shows representative images of the 2D models.
These models consist of particulate media of identical particles, particulate media composed of particles with different sizes (i.e., hyperuniform polydisperse packing), and non-particulate ones (i.e., Debye random medium).
Thus, we take the common characteristic inhomogeneity length scale to be the inverse of the specific surface $1/s$.
Throughout this work, we consider the cases where there are domains of a disconnected phase (phase 2) in a connected matrix phase (phase 1).
We take the volume fractions of phase 2 to be $\phi_2=0.25$ for $d=2$ and $\phi_2=0.125$ for $d=3$.
These values are the largest possible volume fractions that can be achieved across all of the models considered in this paper; see Sec. \ref{sec:step}.

\begin{figure*}
\centering
{\includegraphics[width=0.35\textwidth]{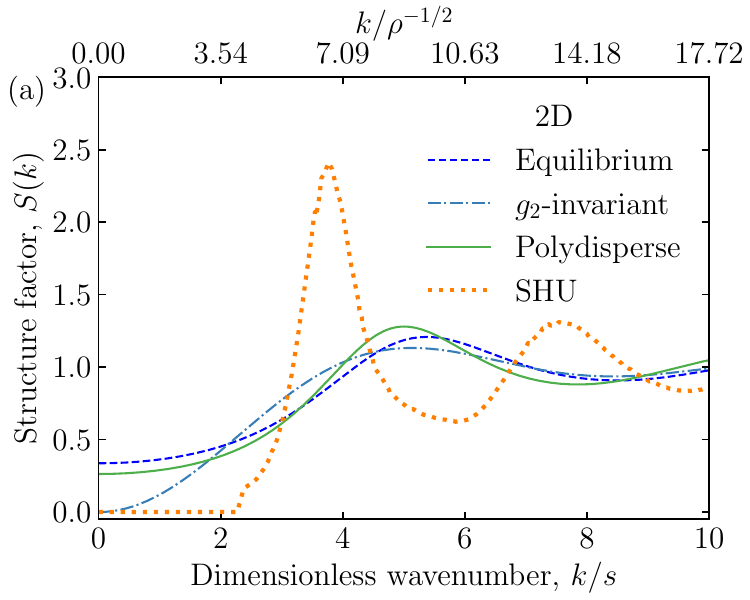}}
{
\includegraphics[width=0.35\textwidth]{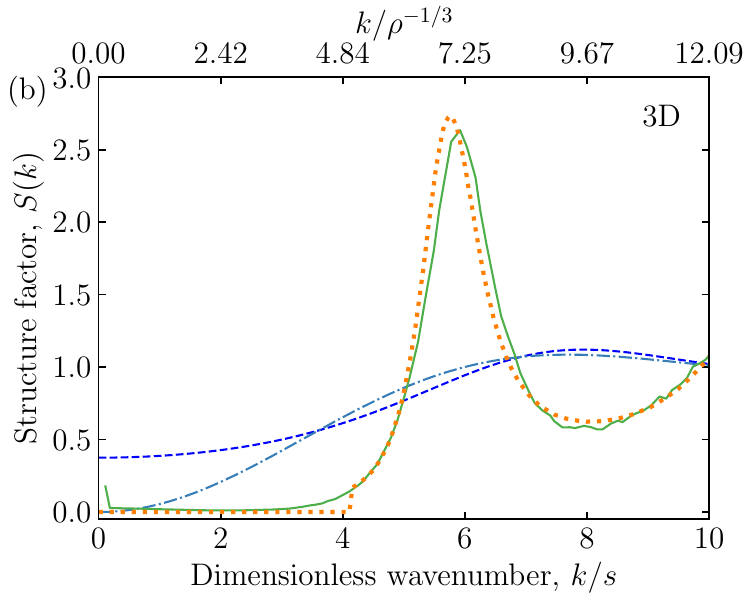}}
\caption{
Structure factor $S(k)$ as a function of the dimensionless wavenumber $k/s$ for the four packing models of 2D two-phase media with $\phi_2=0.25$ (a) and 3D two-phase media with $\phi_2=0.125$ (b).
The upper $x$ axis in each panel represents another dimensionless wavenumber $k/\rho^{-1/d}$ for three packing models: equilibrium, $g_2$-invariant, and SHU packings.
Importantly, the Debye model is not shown here because it is a non-particulate medium.
\label{fig:Sk}
}
\end{figure*}

\begin{figure*}%[b]
\centering
{\includegraphics[width=0.35\textwidth]{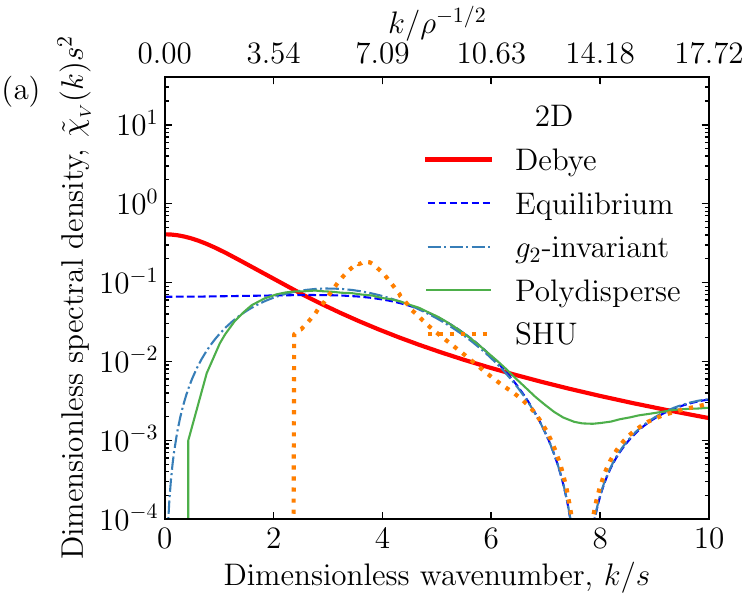}}
{
\includegraphics[width=0.35\textwidth]{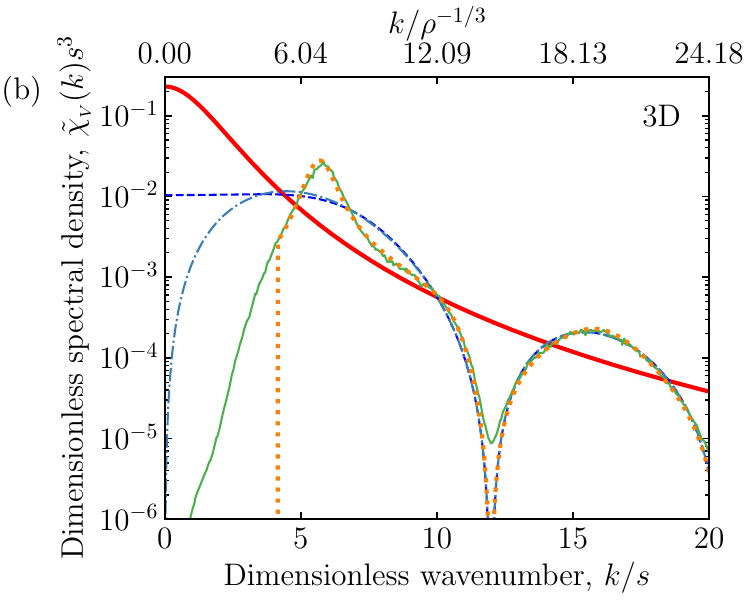}}
\caption{
 Semi-log plots of the dimensionless spectral density $\tilde{\chi}_{_V}(k)s^d$ as a function of the dimensionless wavenumber $k/s$ for the five models of 2D two-phase media with $\phi_2=0.25$ (a) and 3D media with $\phi_2=0.125$ (b).
 The upper $x$ axis in each panel represents another dimensionless wavenumber $k/\rho^{-1/d}$ for three packing models: equilibrium, $g_2$-invariant, and SHU packings.
\label{fig:chik}
}
\end{figure*}

For all models, we compute the structure factor (see Fig. \ref{fig:Sk}) and the spectral density (see Fig. \ref{fig:chik}). 
Using them, we evaluate the scattering mean free paths $\ell_s$ in Sec. \ref{sec:results}.
Such structural features of models are summarized in Table \ref{tab:summary-model} of Appendix \ref{app:models}.
For three selected models, we employ the configurations of models in FDTD simulations to extract their scattering mean free paths (see Sec. \ref{sec:results} and \SI).

\subsection{Debye Random Media}\label{sec:debye}

{\it Debye random media} are characterized by the following form of $\chi_{_{V}}(r)$\cite{yeong_reconstructing_1998,torquato_random_2002}
\begin{align} \label{eq:auto-Debye}
\chi_{_V}(r)=\phi_1\phi_2 \exp(-r/a),
\end{align}
where $a$ is a characteristic length scale. 
Debye {\it et al.} \cite{debye_scattering_1957} proposed that this autocovariance function describes isotropic random media in which the phases create domains of ``random shapes and sizes."
This functional form for $\chi_{_V}(r)$ was later shown to be realizable by two-phase media across space dimensions \cite{skolnick_understanding_2021,torquato_predicting_2020, yeong_reconstructing_1998}.
Specifically, the image of the microstructure shown in Fig. \ref{fig:model}(a) is reproduced from the data published in Ref. \cite{skolnick_understanding_2021}.

For any space dimension $d$, the exact expression for the specific surface\cite{torquato_random_2002} and the spectral density\cite{torquato_predicting_2020} are known; see Appendix \ref{app:Debye}.
We use them to evaluate $\tilde{\chi}_{_V}(k)$ and other properties.

\subsection{Equilibrium Packings}\label{sec:ehd}

Another disordered nonhyperuniform model we treat is the equilibrium (Gibbs) distribution of identical hard spheres of radius $a$ along the stable fluid branch \cite{hansen_theory_1990, torquato_random_2002}.
The structure factors of such packings are well approximated by some (semi-)analytic expressions. \cite{hansen_theory_1990, torquato_random_2002, guo_theoretical_2006}
We use the semianalytic approximation by Guo and Riebel\cite{guo_theoretical_2006} for $d=2$ and the analytic Percus-Yevick solution;\cite{hansen_theory_1990, torquato_random_2002} see Appendix \ref{app:equilibrium}.
Using these approximations in conjunction with \eqref{eq:chi_V-S} yields the corresponding spectral density ${\tilde \chi}_{_V}(k)$.
We also use 2D and 3D packings, generated from the Monte Carlo method,\cite{torquato_random_2002} to perform FDTD simulations for $d=2,3$ and in Fig. \ref{fig:model}(b).

\subsection{Hyperuniform $g_2$-Invariant Sphere Packing}\label{sec:step}

The $g_2$-{\it invariant process} introduced by Torquato and Stillinger\cite{torquato_controlling_2002} aims to determine the range of number density $\rho$ over which a prescribed form of $g_2(r)$ for a many-body system remains invariant. 
They showed that this process has an upper terminal density $\rho_c$. 
In the case of the step-function $g_2$, that is,
\begin{align} 
g_2(r) = \Theta(r-D) =
\begin{cases} 
0, & r\leq D\\
1, & r> D ,
\end{cases}
\end{align}
which corresponds to a packing of identical spheres of diameter $D$, if realizable, up to the terminal packing fraction is $\phi_c \equiv \rho_c v_1(D/2)=2^{-d}$ for any $d$.
At this terminal packing fraction, the structure factor $S(k)$ is known exactly (see Appendix \ref{app:g2-inv}), corresponding to a packing that is hyperuniform of class I with a hyperuniformity exponent $\alpha=2$.\cite{torquato_local_2003}
The realizability of such sphere packings for packing fractions up to the terminal value $\phi_c=1/2^d$ has been confirmed numerically for $d=2,3$ using certain long-ranged effective pair interactions.\cite{wang_equilibrium_2023}
We use Eq. \eqref{eq:step-Sofk-terminal} to compute $S(k)$ and $\tilde{\chi}_{_V}(k)$ in Figs. \ref{fig:Sk} and \ref{fig:chik}.
Some 2D and 3D configurations created via the effective potential approach are used in FDTD simulations and in Fig. \ref{fig:model}(c); see Appendix \ref{app:g2-inv}.

\subsection{Hyperuniform Polydisperse Sphere Packings} \label{sec:NSHU}

We create 2D and 3D hyperuniform sphere packings with a polydispersity in size in a matrix from nonhyperuniform progenitor point patterns via a tessellation-based procedure;\cite{kim_new_2019, kim_methodology_2019} see details in Appendix \ref{app:NSHU}.
These configurations are used to create Fig. \ref{fig:model}(d) and to compute $S(k)$ and $\tilde{\chi}_{_V}(k)$ numerically in Fig. \ref{fig:Sk}, and are also employed in FDTD simulations for $d=2,3$.
In the thermodynamic limit, the spectral densities of the resulting packings exhibit a power-law scaling $\tilde{\chi}_{_V}(\vect{k}) \sim \abs{\vect{k}}^4$ for small wavenumbers.\cite{kim_methodology_2019}
The simulation parameters used to generate the representative statistical ensembles are $\phi_b=0.30$ and $N=100$ for $d=2$, and $\phi_b=0.45$ and $N=100000$ for $d=3$.
For FDTD simulations in two and three dimensions, we use configurations with $N=100$ and $N=1000$, respectively.

\subsection{SHU Sphere Packings} \label{sec:SHU}

SHU two-phase media have $\tilde{\chi}_{_V}(\vect{k})=0$ for the finite range $0<\left\vert \vect{k}\right\vert\leq K$, called the {\it exclusion region}.
We specifically consider $d$-dimensional SHU sphere packings in a matrix with packing fraction $\phi_2$.
For such SHU media, the degree of stealthiness $\chi$ is measured by the ratio of the number of the wave vectors within the exclusion region in the Fourier space to the total degrees of freedom, that is, $\chi = v_1(K)/[2d (2\pi)^d \rho ]$.\cite{torquato_ensemble_2015}
These SHU systems are highly degenerate and disordered if $\chi <1/3$ in one dimension or $\chi<1/2$ in two and three dimensions \cite{zhang_ground_2015}.
Thus, we numerically create 2D and 3D packings for $\chi=0.35$, and their respective packing fractions are $\phi_2=0.25,0.125$; see details in Appendix \ref{app:SHU}.
These configurations are used to create Fig. \ref{fig:model}(e) and to compute $S(k)$ and $\tilde{\chi}_{_V}(k)$ in Fig. \ref{fig:Sk}, and are also employed in FDTD simulations for $d=2,3$.

\begin{figure*}[t!]
 {
 \includegraphics[width=0.45\textwidth]{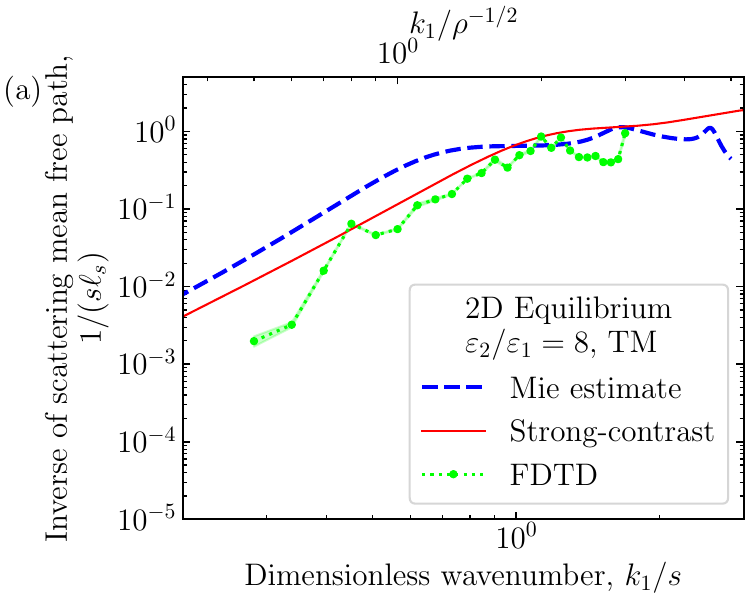}}
 {
 \includegraphics[width=0.45\textwidth]{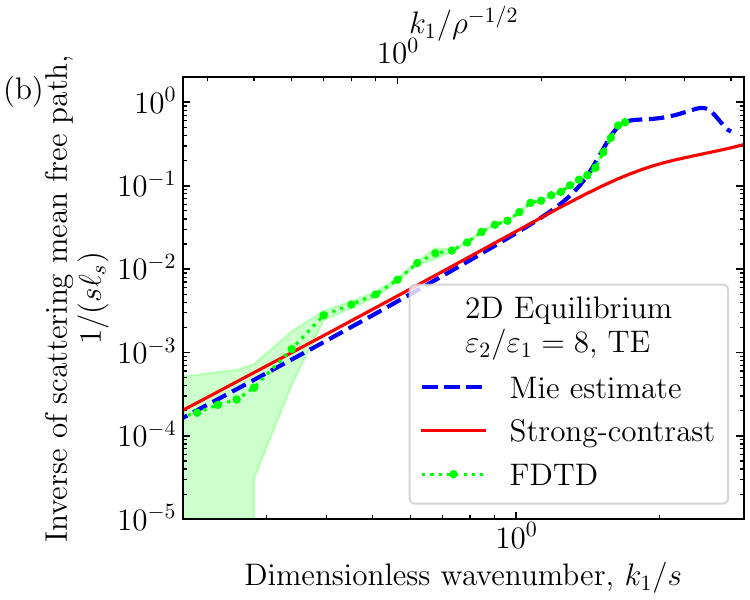}}
\caption{
Log-log plots of the inverse of dimensionless scattering mean free path $1/(s\ell_s)$ as a function of the dimensionless incident wavenumber $k_1/s$ for 2D equilibrium packings with $\phi_2=0.25$ and $\varepsilon_2/\varepsilon_1=8$ for TM (a) and TE (b) polarizations.
The upper $x$ axis in each panel represents another dimensionless wavenumber $k/\rho^{-1/2}$.
The estimations from FDTD simulations, the Mie estimate given in Eq. \eqref{eq:Mie-ells}, and the strong-contrast estimates using Eq. \eqref{eq:ells-effective-medium}.
% We choose the lower contrast ratio $\varepsilon_2/\varepsilon_1=4$ for TM polarization than for TE polarization because of the convergence issues in FDTD simulations.
The green shades represent the statistical errors in FDTD simulations.
\label{fig:2D_EP_FDTD}
 }
\end{figure*}

\subsection{Structure Factors and Spectral Densities}
\label{sec:Sk-chik}

Estimation of $\ell_s$ using the Mie estimate \eqref{eq:Mie-ells} and the strong-contrast approximations in Eq. \eqref{eq:ells-effective-medium} requires the computation of the structure factor $S(k)$ and spectral density $\tilde{\chi}_{_V}(k)$, respectively.
Figure \ref{fig:Sk} shows $S(k)$ of the four packing models (i.e., except for the Debye random media) as a function of the dimensionless wavenumber $k_1/s$.
In Fig. \ref{fig:Sk} and the ensuing plots, we add secondary upper $x$ axes to aid comparison with a commonly used scale $k_1/\rho^{-1/d}$.
For a packing of identical disks/spheres of radius $a$ and packing fraction $\phi_2$, the two length scales are related by $\rho^{-1/d}=s / [d \phi_2 v_1(a)]$.
Among them, $S(k)$ of the equilibrium packings and $g_2$-invariant packings are evaluated from the analytic expressions, whereas those of hyperuniform polydisperse and SHU packings are evaluated from numerically generated configurations.
Figure \ref{fig:chik} shows $\tilde{\chi}_{_V}(k)$ of all models, in which the curves for hyperuniform polydisperse and SHU packings are evaluated from numerically generated configurations, whereas those of the other models are evaluated from the analytic formulas.

Importantly, the spectral density is more versatile than the structure factor when it comes to characterizing general two-phase media.
First, for non-particulate media, such as Debye random media, the structure factor cannot be evaluated.
Second, for more general particulate media, like hyperuniform polydisperse packings, the structure factor cannot account for the variations in particle sizes.
Thus, while $S(k)$ in Fig. \ref{fig:Sk} indicates that this polydisperse packing is nonhyperuniform [i.e., $S(0)>0$], $\tilde{\chi}_{_V}(k)$ in Fig. \ref{fig:chik} indicates that this model is indeed hyperuniform.
We show that the latter leads to more accurate estimates in Sec. \ref{sec:FDTD}.
Because the spectral densities $\tilde{\chi}_{_V}(k)$ also contain particle-shape information via the term $\tilde{m}(k;a)$ [see Eqs. \eqref{eq:chi_V-S} and \eqref{eq:spd-packing}], $\tilde{\chi}_{_V}(k)$ of the particulate media of identical particles collapse onto the same curves for large wavenumbers, that is, $k_1/s \gtrsim 8$ for $d=2,3$.

\section{Results}\label{sec:results}

Here, we present the strong-contrast estimates for $1/(s\ell_s)$, for both 2D and 3D models explained in Sec. \ref{sec:models}, that are evaluated by substituting Eqs. \eqref{eq:trans-TM}, \eqref{eq:trans-TE}, and \eqref{eq:iso-2pt}, into Eq. \eqref{eq:ells-effective-medium}.
However, we do not present the results for $d=1$ here because their qualitative behaviors are similar to those for higher dimensions.
The attenuation functions, evaluated in an intermediate step, are shown in Appendix \ref{app:3D_F}.
We begin by verifying the accuracy of these strong-contrast estimates, comparing them to those from the Mie estimate \eqref{eq:Mie-ells} and FDTD simulations for three selected 2D models, i.e., equilibrium packing, hyperuniform $g_2$-invariant packings, and hyperuniform polydisperse packings; see Sec. \ref{sec:FDTD}.
Additional FDTD simulation results--(i) larger 2D realizations of the three models at $\phi_2=0.25$, (ii) 2D equilibrium and SHU models at higher packing fractions, and (iii) the three 3D models at $\phi_2=0.125$--are provided in the \SI.
We then compare the behaviors of $\ell_s$ from the strong-contrast estimates across models for $d=2,3$; see Sec. \ref{sec:comp}.

\subsection{Numerical Verifications via the FDTD Simulations} 
\label{sec:FDTD}

Here, we corroborate the accuracy of the strong-contrast estimates for $\ell_s$ (Sec. \ref{sec:SCA}) by comparing them with FDTD simulations for 2D models with $\phi_2=0.25$ for both TE and TM polarizations.
We choose a relatively high dielectric contrast ratio of $\varepsilon_{2}/\varepsilon_{1}=8$ for the reasons given in Sec. \ref{sec:intro}.
We also compare our strong-contrast predictions of the scattering mean free path to the Mie estimate \eqref{eq:Mie-ells}, when applicable.
We consider four selected models of particulate media: equilibrium packing, hyperuniform $g_2$-invariant packing, hyperuniform polydisperse packing, and SHU packings.
Equilibrium packings serve as a common benchmark, since the Mie estimate \eqref{eq:Mie-ells} is known to be accurate for this model at moderate packing fractions ($\phi_2\lesssim 0.4$) in two\cite{conley_light_2014,riboli_tailoring_2017} and three\cite{fraden_multiple_1990,saulnier_scatterer_1990} dimensions.
Results for 2D equilibrium packings at $\phi_2>0.25$ are also computed and reported in the \SI.
We perform FDTD simulations for two additional hyperuniform models, where the Mie estimate does not apply to the polydisperse packing but applies to the $g_2$-invariant one.
The analogous results for SHU packings with $\phi_2=0.25,0.40$ are presented in the \SI~ for the sake of brevity.
However, FDTD simulations are not conducted for the Debye media with $\phi_2=0.25$ because they exhibit two microstructural features: (i) nonhyperuniformity and (ii) packings of particles with varying sizes and shapes, as shown in Fig. \ref{fig:model}(a) and a rigorous mapping to overlapping spheres.\cite{skolnick_understanding_2021}
We validate strong-contrast estimates for these two structural features using the equilibrium packing and the polydisperse packing.

We also corroborate the accuracy of the strong-contrast estimates for $\ell_s$ in three dimensions ($d=3$) by comparing them with FDTD simulations for three models: equilibrium packing, hyperuniform $g_2$-invariant packing, and hyperuniform polydisperse packing; see \SI.
We consider solely the cases of $\phi_2=0.125$.

\begin{figure*}
 {
 \includegraphics[width=0.45\textwidth]{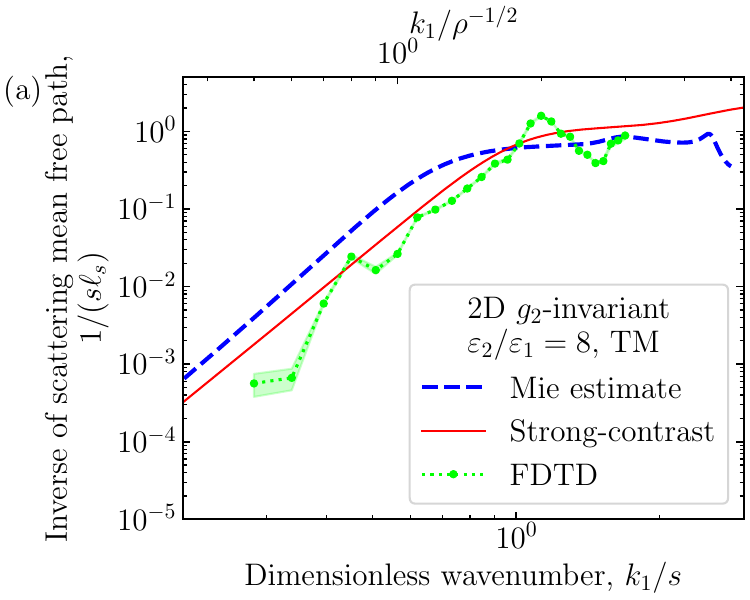}}
 {
 \includegraphics[width=0.45\textwidth]{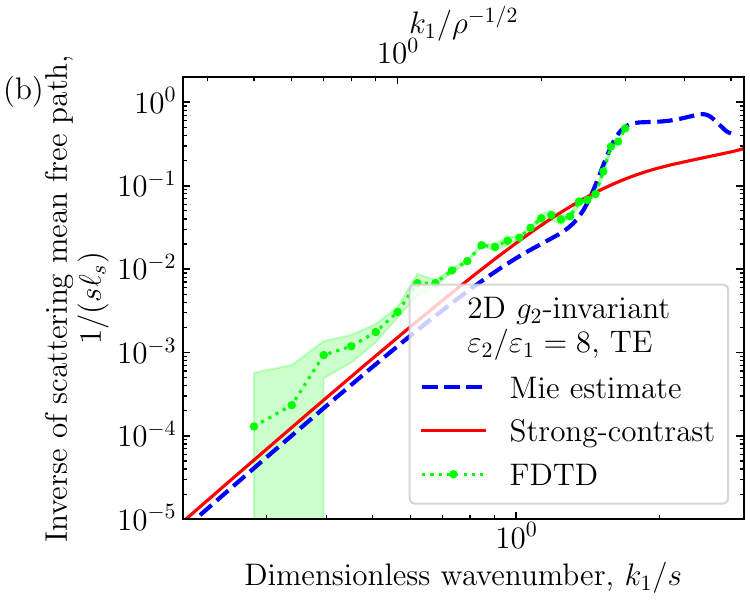}}
\caption{Log-log plots of the inverse of dimensionless scattering mean free path $1/(s\ell_s)$ as a function of the dimensionless incident wavenumber $k_1/s$ for 2D hyperuniform $g_2$-invariant sphere packings with $\phi_2=0.25$ and $\varepsilon_2/\varepsilon_1=8$ for TM (a) and TE (b) polarizations.
The upper $x$ axis in each panel represents another dimensionless wavenumber $k/\rho^{-1/2}$.
The estimations from FDTD simulations, the Mie estimate \eqref{eq:Mie-ells}, and the strong-contrast estimates using Eq. \eqref{eq:ells-effective-medium}.
The green shades represent the statistical errors in FDTD simulations.
 \label{fig:2D_g2_FDTD}
 }
\end{figure*}

\begin{figure*}
 {
 \includegraphics[width=0.45\textwidth]{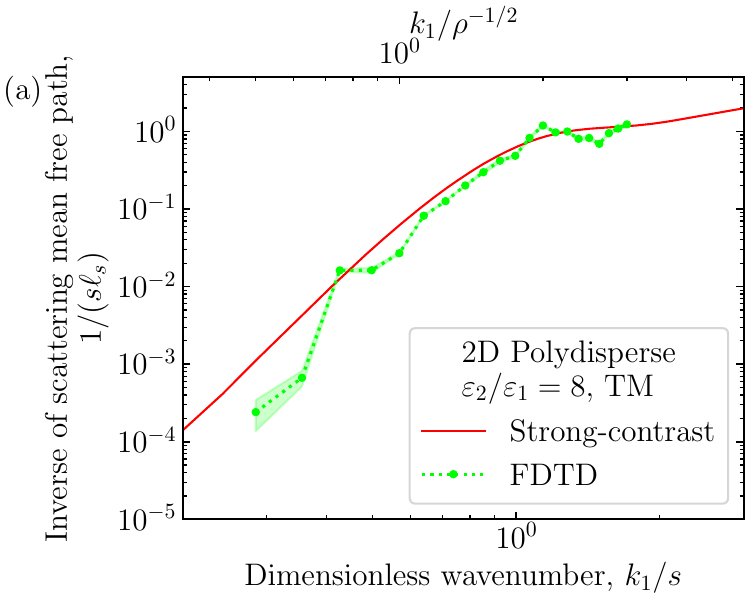}}
 {
 \includegraphics[width=0.45\textwidth]{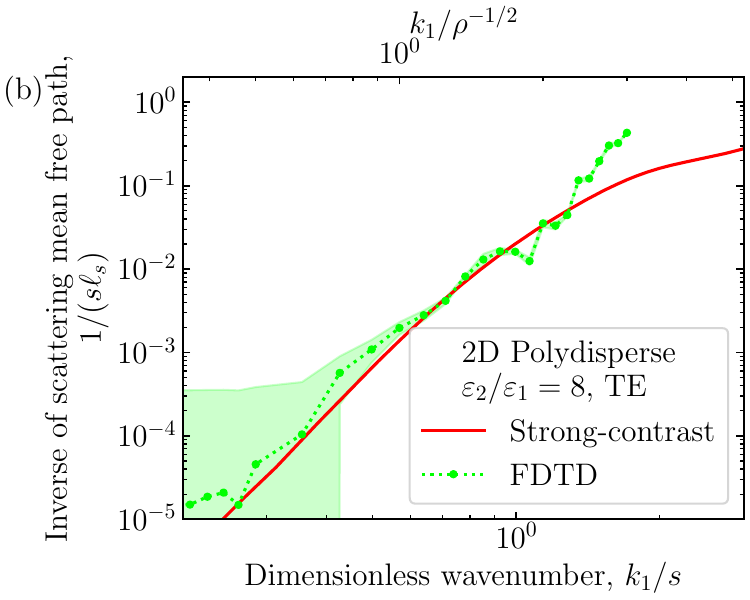}}
\caption{Log-log plots of the inverse of dimensionless scattering mean free path $1/(s\ell_s)$ as a function of the dimensionless incident wavenumber $k_1/s$ for 2D hyperuniform polydisperse sphere packings with $\phi_2=0.25$ and $\varepsilon_2/\varepsilon_1=8$ for TM (a) and TE (b) polarizations.
The upper $x$ axis in each panel represents another dimensionless wavenumber $k/\rho^{-1/2}$.
The estimations from FDTD simulations and the strong-contrast estimates using Eq. \eqref{eq:ells-effective-medium}.
The Mie estimate \eqref{eq:Mie-ells} is not shown here because it is inapplicable to this model.
The green shades represent the statistical errors in FDTD simulations.
\label{fig:2D_poly_FDTD}
 }
\end{figure*}

In FDTD simulations, we estimate the scattering mean free path $\ell_s$ of a model of random media from the ballistic Poynting vector\cite{jackson_classical_1999} 
\begin{align} \label{eq:S_b}
S_{x,\mathrm{ballistic}}\equiv \mathrm{Re}[\expval{\vect{E}}\times \expval{\vect{H}}^*]/2.
\end{align}
In Eq. \eqref{eq:S_b}, the coherent parts of the electric and magnetic fields, denoted by $\expval{\vect{E}}$ and $\expval{\vect{H}}$ respectively, are obtained by averaging the fields over numerous realizations of the media. 
Such coherent fields can be used to estimate the effective dielectric constants.\cite{pattelli_role_2018, kim_theoretical_2024}
To improve numerical convergence, we also perform a spatial average in the direction perpendicular to the propagation direction. Subsequently, we fit the average intensity using the following formula\cite{monsarrat_pseudogap_2022, vynck_light_2023}
\begin{align} \label{eq:S_fit}
 S_{x,\mathrm{ballistic}} = A \exp(-x/\ell_s),
\end{align}
where $x$ is the propagation distance from the boundary of a medium.
In contrast to Refs. \cite{monsarrat_pseudogap_2022,vynck_light_2023}, we choose not to use $\abs{\expval{\vect{E}}}^2$ for our analysis because this quantity exhibits strong oscillations along $x$ for long wavelengths, which prevent accurate fitting; see details in Appendix \ref{app:FDTD}.
We remark that the estimates obtained from FDTD simulations tend to be less accurate (as indicated by the broader green-shaded error bars in Figs. \ref{fig:2D_EP_FDTD}, \ref{fig:2D_g2_FDTD}, and \ref{fig:2D_poly_FDTD}), when the wavenumber decreases.
This loss of numerical accuracy originates from the exponential fit in Eq. \eqref{eq:S_fit} at a fixed system size $L$, which becomes less accurate when $\ell_s \gg L$ (i.e., in the regime of small optical thickness) where the uncertainties in the electric and magnetic fields are comparable to the scattering attenuation across the system.
For this reason, the error starts to increase as $1/(s\ell_s)$ becomes lower than $\delta E/(EL) \sim 10^{-2} - 10^{-3}$, where $\delta E / E \sim 0.05$ represents the relative statistical error in the averaged electric field. 
These errors diminish as $L$ increases (see Figs. S1 and S2), and the larger configurations ($N=400$) yield results consistent with the smaller ones ($N=100$) at $\phi_2=0.25$.

\begin{figure*}[th!]
\centering
 {\includegraphics[width=0.4\textwidth]{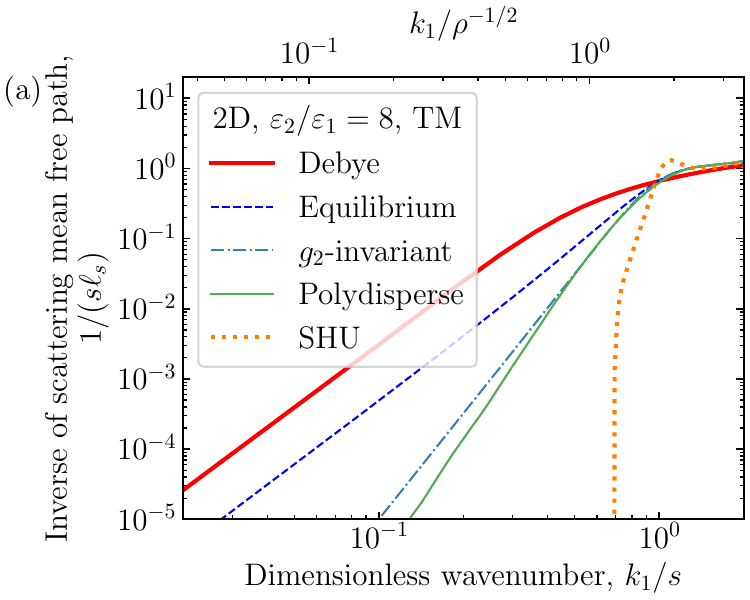}}
 {\includegraphics[width=0.4\textwidth]{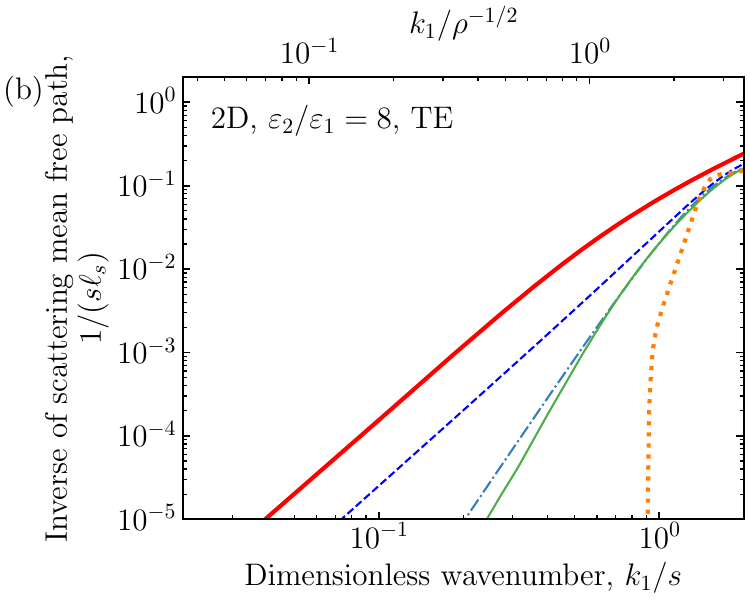}}
 \caption{
 Log-log plots of the inverse of dimensionless scattering mean free path $1/(s\ell_s)$, obtained from the strong-contrast estimates [i.e., substitution of Eqs. \eqref{eq:trans-TM} and \eqref{eq:trans-TE} in Eq. \eqref{eq:ells-effective-medium}], as a function of the dimensionless incident wavenumber $k_1/s$ for all 2D models with $\phi_2=0.25$ and $\varepsilon_2/\varepsilon_1=8$.
 Panels (a) and (b) are for TM and TE polarizations, respectively.
 The upper $x$ axis in each panel represents another dimensionless wavenumber $k/\rho^{-1/2}$ for three packing models: equilibrium, $g_2$-invariant, and SHU packings. 
 These values are slightly smaller than those found for the hyperuniform polydisperse packing.
 \label{fig:2D_all}
 }
\end{figure*}

We examine 2D equilibrium packings, where the Mie estimate \eqref{eq:Mie-ells} is also applied; see Fig. \ref{fig:2D_EP_FDTD}.
We consider dielectric constant ratio $\varepsilon_2/\varepsilon_1=8$ for both TM and TE polarizations.
We note that both the Mie estimate \eqref{eq:Mie-ells} and the strong-contrast approximations in Sec. \ref{sec:SCA} as well as the simulations reveal that the inverse of scattering mean free path $1/\ell_s$ increases up to the intermediate-wavenumber regime $k_1/s\lesssim1$.
This behavior occurs for all models because the scattering strength associated with the inhomogeneous dielectric constant increases with the wavenumber in this regime.
In the equilibrium-packing model, the Mie estimate \eqref{eq:Mie-ells} and the strong-contrast estimate in Sec. \ref{sec:SCA} provide similar results, and both estimates are in excellent agreement with FDTD simulations for $ k_1/s \lesssim 1$.
Within this range, however, the strong-contrast approximation is more accurate for TM polarization.
Beyond this range (i.e., $k_1/s\gtrsim 1$), however, since the contribution from scattering of a single particle becomes increasingly important, the Mie estimate \eqref{eq:Mie-ells} becomes more accurate than the corresponding strong-contrast approximation.
This difference is particularly pronounced for TE polarization, where $\dv*{\sigma_s}{\Omega}$ in the Mie theory can vary significantly in its angular dependence compared to the particle form factor used in the strong-contrast approximation; see discussion in Sec. \ref{sec:long-wavelength} and Appendix \ref{app:FormFactor-Mie}.

We make an analogous comparison for the $g_2$-invariant packings, where the Mie estimate is still applicable; see Fig. \ref{fig:2D_g2_FDTD}.
For both TM and TE polarizations, we take $\varepsilon_2/\varepsilon_1=8$.
The strong-contrast approximations also show good agreement with the simulations up to $k_1/s\lesssim 1$ for this model.
Within the same range, the strong-contrast approximation is more accurate than the Mie estimate for TM polarization but yields similar results for TE polarization, as seen in the equilibrium-packing case (Fig. \ref{fig:2D_EP_FDTD}).
The Mie estimate still captures a sudden increase in $1/\ell_s$ around $k_1/s\approx 1$ for TE polarization, as in Fig. \ref{fig:2D_EP_FDTD}.
These discrepancies suggest that while the Mie estimate accurately accounts for scattering by a single particle, it does not accurately capture the long-ranged many-particle spatial correlations exhibited in hyperuniform media.

Figure \ref{fig:2D_poly_FDTD} compares the strong-contrast approximations and the simulations for hyperuniform polydisperse packings with $\varepsilon_2/\varepsilon_1=8$, where the Mie estimate is no longer applicable.
Similar to the two previous models, the strong-contrast approximations also show good agreement with the simulations up to $k_1/s\lesssim 1$ for both TM and TE polarizations.

For the same 2D models studied in Figs. \ref{fig:2D_EP_FDTD}-\ref{fig:2D_poly_FDTD}, we also perform FDTD simulations for larger configurations with $N=400$ particles to examine finite-size effects; see the \SI.
We observe that this larger system size diminishes the statistical errors in $1/\ell_s$ for small wavenumbers (as explained earlier), whereas we do not observe appreciably sized systematic errors for $k_1/s \lesssim 1$.

We also test the accuracy of the strong-contrast estimates at packing fractions greater than or equal to $\phi_2=0.25$ for 2D equilibrium packings ($\phi_2=0.40, 0.50$) and SHU packings ($\phi_2=0.25,0.40$); see Figs. S3-S4 in the \SI.
For both equilibrium and SHU packings, the trends are similar to those at $\phi_2=0.25$ (see Fig. \ref{fig:2D_EP_FDTD}).
For $k_1/s \lesssim 1$, the Mie and strong-contrast estimates for $1/\ell_s$ remain close to the FDTD results;
for $k_1/s \gtrsim 1$, particularly for TE polarization, the Mie estimate is more accurate, although this advantage weakens as $\phi_2 \ge 0.40$.

We also numerically verify our 3D predictions by conducting FDTD simulations with some models at $\phi_2=0.125$; see Fig. S5 in the \SI.
We consider equilibrium packings, hyperuniform $g_2$-invariant packings, and hyperuniform polydisperse packings for transverse polarization with $\varepsilon_2/\varepsilon_1=8$ and $N=1000$.
We observe that the strong-contrast and Mie estimates are nearly the same, and both agree well with FDTD results for $k_1/s \lesssim 2$.
Beyond this range (i.e., $k_1/s \gtrsim 2$), the Mie estimate becomes slightly more accurate than the strong-contrast approximations.
These trends are similar to those for 2D models shown in Figs. \ref{fig:2D_EP_FDTD}-\ref{fig:2D_poly_FDTD}, except for the slightly different crossover condition, that is, $k_1/s\approx 2$.
While we do not perform simulations at higher packing fractions here, the accuracy of strong-contrast approximations for effective dielectric constants has been established for equilibrium packings and SHU packings at $\phi_2=0.25$.\cite{torquato_nonlocal_2021}

\subsection{Predictions for Scattering Mean Free Paths Across the Different Models}
\label{sec:comp}

Having established the accuracy of the strong-contrast estimates for $\ell_s$ in the range of $k_1/s \lesssim 1$ (see Sec. \ref{sec:FDTD}), we now compare the predictions of Eqs. \eqref{eq:trans-TM}, \eqref{eq:trans-TE}, and \eqref{eq:iso-2pt}, in conjunction with Eq. \eqref{eq:ells-effective-medium}, for all five different models in two and three dimensions.
For simplicity, we focus on the cases with a fixed contrast ratio $\varepsilon_2/\varepsilon_1=8$.

We begin by describing the 2D results presented in Fig. \ref{fig:2D_all}. For TM polarization, shown in Fig. \ref{fig:2D_all}(a), the relative ranking in the degree of attenuation, measured by $1/(s\ell_s)$, remains consistent for a wide range of wavenumbers up to $k_1/s \approx 0.9$. 
Specifically, the two nonhyperuniform models exhibit significant degrees of attenuation, with the Debye medium showing the highest values of $1/(s\ell_s)$, followed by the equilibrium packing, and their scaling behavior is the same. 
In contrast to the nonhyperuniform cases, the three hyperuniform media exhibit appreciably less attenuation. Their scaling behavior of $1/\ell_s$ varies widely: it follows $1/\ell_s \sim {k_1}^{5}$ for the $g_2$-invariant packing, $1/\ell_s \sim {k_1}^{7}$ for the polydisperse packing, and $1/\ell_s \sim 0$ for the SHU ones, indicating perfect transparency.

For the same 2D models, the corresponding microstructure-dependent behaviors for $1/\ell_s$ also occur for TE polarization, as shown in 
Fig. \ref{fig:2D_all}(b). 
One difference is that the same model always exhibits a higher degree of attenuation for TM polarization than for TE polarization at a given wavenumber $k_1$. 
This observation is consistent with the well-known result\cite{vandehulst_light_1957} that the Rayleigh scattering cross-section of a cylindrical particle is larger for TM polarization than for TE polarization by a factor of $(\varepsilon_2+\varepsilon_1)^2/2$.

\begin{figure}
\centering
 \includegraphics[width=0.4\textwidth]{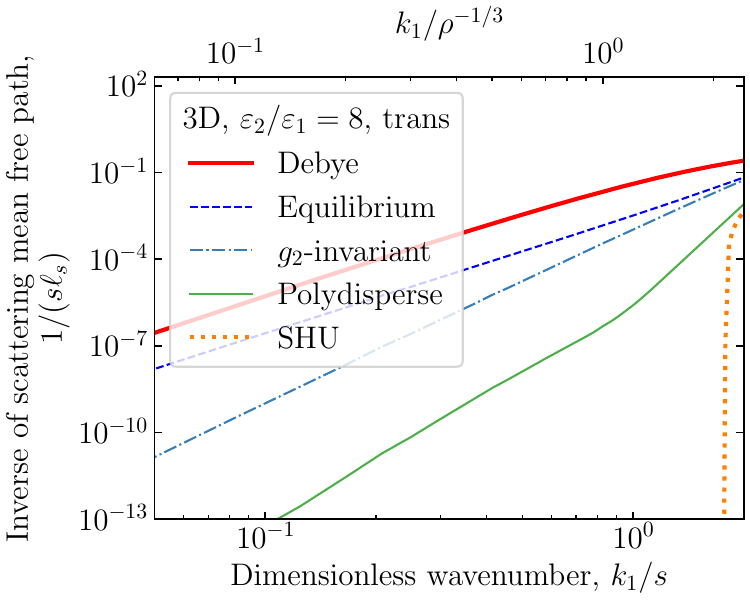}
 \caption{
 Log-log plots of the inverse of dimensionless scattering mean free path $1/(s\ell_s)$, predicted from the strong-contrast estimate using Eq. \eqref{eq:iso-2pt}, as a function of the dimensionless incident wavenumber $k_1/s$ for all 3D models with $\phi_2=0.125$, $\varepsilon_2/\varepsilon_1=8$.
 The upper $x$ axis represents another dimensionless wavenumber $k/\rho^{-1/3}$ for three models: equilibrium, $g_2$-invariant, and SHU packings.
 Estimates for transverse polarization are considered.
 \label{fig:3D_all}
 }
\end{figure}

Figure \ref{fig:3D_all} makes comparisons of $1/\ell_s$ for 3D (fully statistically isotropic) models with $\phi_2=0.125$ for transverse polarization. The qualitative behaviors for $d=3$ are largely similar to the 2D models aforementioned in Fig. \ref{fig:2D_all}. 
The Debye medium and the equilibrium packing exhibit the same scaling with $1/\ell_s \sim k_1^{4}$ with the highest degrees of attenuation. 
Similar to the 2D cases, the three 3D hyperuniform models also exhibit a wide range of scaling behavior of $1/\ell_s$, ranging from $1/\ell_s \sim k_1^{6}$ for the $g_2$-invariant packing to perfect transparency for the SHU ones.
Notably, the scattering behavior of the hyperuniform polydisperse packings, characterized by $1/\ell_s \sim k_1^{8}$, is consistent with the anomalous light scattering observed in nanostructured glasses, which Shepilov\cite{shepilov_anomalous_2023} modeled using the same microstructural model and attributed to the interference of Rayleigh scatterers of different sizes.

\section{Discussion}\label{sec:conclusions}

The present work provides new predictive formulas for the scattering mean free path $\ell_s$ of disordered 1D (layered), 2D (transversely isotropic), and 3D (fully statistically isotropic) two-phase media, derived from the strong-contrast approximations for the effective dielectric constant; see Sec. \ref{sec:SCA}.
Importantly, these new formulas are applicable to general disordered two-phase media, including monodisperse sphere packings, polydisperse ones, and even non-particulate media.
For three selected 2D models and dielectric constant ratio $\varepsilon_2/\varepsilon_1=8$, we have validated that these formulas are accurate beyond the long-wavelength regime $k_1/s \lesssim 1$ by comparing them with FDTD simulations; see Sec. \ref{sec:FDTD}.
In the same range $k_1/s\lesssim 1$, these predictions are as accurate as the Mie estimates \eqref{eq:Mie-ells} for TE polarization, but they are more accurate for TM polarization.
For $k_1/s \gtrsim 1$, however, the Mie estimate captures the behavior of $1/\ell_s$ more accurately, as expected.
These observations align with the physical interpretation that while the strong-contrast approximation partially accounts for near-field effects omitted by the Mie estimate, it lacks the Mie estimate's ability to fully resolve particle-specific modes, as elaborated in Sec. \ref{sec:SCA}.
It is therefore instructive to relate this crossover condition, $k_1/s \sim 1$, to the corresponding physical wavelength and particle size.
For monodisperse sphere packings in $d$ dimensions, the crossover condition translates to the diameter-to-wavelength ratio, that is, $2a/\lambda \sim d\phi_2/\pi$ is around $0.5$ or smaller.
More specifically, in rutile $\mathrm{TiO}_{2}$/air systems at wavelengths near 700~nm ($\varepsilon_2/\varepsilon_1=8$),\cite{polyanskiy_refractiveindexinfo_2024} this crossover condition implies a particle diameter of approximately $2a \approx 110$~nm for two-dimensional disk packings with $\phi_2=0.25$.
While we conducted FDTD simulations mainly for 2D model microstructures here, we expect that the strong-contrast estimates for $\ell_s$ are accurate for 1D and 3D models as well, given that prior work shows that the underlying approximations for $\varepsilon_\mathrm{e}$ are very accurate for such 1D\cite{kim_effective_2023,kim_theoretical_2024} and 3D media.\cite{torquato_nonlocal_2021}

Having established the accuracy of the strong-contrast estimates of $\ell_s$, we applied them to study the dependence of $\ell_s$ on the microstructures for the five different models for $d=2,3$; see Sec. \ref{sec:comp}.
The small-$k_1$ behaviors of $\ell_s$ for the five models can be summarized by the following power-law scaling behaviors:
\begin{align} \label{eq:scaling-ells}
 \frac{1}{\ell_s} 
 \sim 
 \begin{cases}
 {k_1}^{d+1}, & \text{Debye, Equilibrium}, \\
 {k_1}^{d+3}, & \text{$g_2$-invariant},\\
 {k_1}^{d+5}, & \text{Polydisperse},\\
0, & \text{SHU},\\
 \end{cases}
 \quad k_1/s \ll 1,
\end{align}
where $d=1,2,3$.
Importantly, while all typical nonhyperuniform media show a similar behavior in the effective attenuation, various classes of hyperuniform media can have different scaling behaviors.
As we noted in Sec. \ref{sec:transparency}, the zero values of the inverse of scattering mean free path (i.e., $1/\ell_s$) indicate perfect transparency.

It is instructive to remark on another key wave characteristic of disordered two-phase dielectric media, namely, the {\it transport mean free path} $\ell_t$.
While $\ell_s$ governs the attenuation of the coherent (ballistic) intensity, $\ell_t$ characterizes diffusive wave transport in the multiple-scattering regimes.\cite{ishimaru_wave_1978,sheng_introduction_2006,akkermans_mesoscopic_2007,scheffold_transport_2022}
In general, reducing either $\ell_s$ or $\ell_t$ leads to a decrease in the transmitted intensity through a finite sample, although the underlying physical mechanisms differ. 
The theory developed in the present article describes the coherent field inside the medium, as characterized by ensemble averages of the form $\expval{\vect{E}(\vect{x})}$, whereas $\ell_t$ depends on the angular redistribution of intensity, involving second-order quantities such as $\expval{\vect{E}(\vect{x})\cdot\vect{E}(\vect{x})}$.
A rigorous treatment of $\ell_t$ within the strong-contrast framework remains an open problem and a natural direction for future work. 
Nevertheless, qualitative insights can be inferred from Mie-based analyses to devise a strong-contrast estimate, since $\ell_t$ and $\ell_s$ are related through the scattering anisotropy parameter $g$.\cite{vynck_light_2023} 
In particular, for nonstealthy hyperuniform media, the forward-scattering contribution, quantified by $\tilde{\chi}_{_V}(0)$, is anomalously suppressed relative to the backscattering one at $\tilde{\chi}_{_V}(2k_1)$.
This imbalance suggests a negative $g$, meaning that $\ell_s$ is expected to be comparable to, or larger than $\ell_t$ up to the first peak in the spectral density. In sum, the development of a strong-contrast estimate of the transport mean free path is
an outstanding open problem.

In addition to the scaling laws \eqref{eq:scaling-ells}, our general findings have demonstrated that the quantity $\ell_s$ can be continuously modulated by tailoring the spectral density $\tilde{\chi}_{_V}(k)$ at small wavenumbers for fixed phase dielectric constants, thus offering a clear pathway for the inverse design\cite{torquato_inverse_2009,sherman_inverse_2020} of optical materials with desired scattering properties.
Specific examples of desirable scattering properties for practical applications include maximizing $\ell_s$ for transparent gradient index metamaterials,\cite{zhang_experimental_2019} minimizing it for random lasing media\cite{gayathri_lasing_2023} and efficient brighteners,\cite{haataja_topological_2023, pattelli_role_2018} and achieving well-defined maxima of $\ell_s$ for selective filtering materials.\cite{otanicar_filtering_2016,rothammer_tailored_2021}
Subsequently, two-phase media with such targeted spectral densities $\tilde{\chi}_{_V}(k)$ can be computationally constructed using the Fourier-space construction techniques.\cite{chen_designing_2018,shi_computational_2023,shi_threedimensional_2025} 
These computationally designed media can be readily fabricated via 2D photolithographic\cite{zhao_assembly_2018} and 3D printing techniques.\cite{tumbleston_continuous_2015}

%\backmatter

%\backmatter
\bmsection*{Author Contributions}
Jaeuk Kim and Salvatore Torquato contributed equally to this work.

\bmsection*{Acknowledgements}
J.K. and S.T. were supported by the Army Research Office, accomplished under Cooperative Agreement Number W911NF-22-2-0103. 
J.K. was also supported by the InnoCORE program of the Ministry of Science and ICT(GIST InnoCORE KH0830).
Simulations were performed on computational resources managed and supported by the Princeton Institute for Computational Science and Engineering (PICSciE) and the Korea Institute of Science and Technology Information (KISTI).

% \bmsection*{Financial disclosure}

% None reported.

\bmsection*{Conflicts of Interest}
The authors declare no conflicts of interest.

\bmsection*{Data Availability Statement}
The data and code for this work are partly available at https://doi.org/10.5281/zenodo.18419454. 
Additional data are available from the corresponding author upon reasonable request.

\bmsection*{Supporting Information}
Additional supporting information can be found online in the Supporting Information section.

\appendix

\begin{table*}
    \renewcommand{\arraystretch}{1.5}
    \caption{Summary table of five models of disordered two-phase media. 
    From the leftmost to the rightmost model, we list Debye random media, equilibrium packings, hyperuniform $g_2$-invariant packings, hyperuniform polydisperse packings, and SHU packings.
    It indicates applicability of the predictive formulas for $\ell_s$: the Mie estimate \eqref{eq:Mie-ells} or the strong-contrast approximations in Sec. \ref{sec:SCA}.
    \label{tab:summary-model}}
    \begin{tabular}{c | c c c c c}
    \hline
    Model & Debye & Equilibrium packing & $g_2$-invariant & Polydisperse & SHU \\
    \hline
    Hyperuniformity & Non-hyperuniform & Non-hyperuniform & Hyperuniform & Hyperuniform & Stealthy Hyperuniform \\
    Particulate Medium& No & Yes & Yes & Yes & Yes \\
    Particle sizes& Not applicable & Identical & Identical & Different & Identical \\
    Structure factor& Not applicable & Analytic (approx.) & Analytic (exact) & Numerical & Numerical \\
    Spectral density& Analytic (exact) & Analytic (approx.) & Analytic (exact) & Numerical & Numerical \\
    \hline
    Mie estimate& Not applicable & Applicable & Applicable & Not applicable & Applicable \\  
    Strong-contrast& Applicable & Applicable & Applicable & Applicable & Applicable  \\
    \hline
\end{tabular}
\end{table*}

\section{Structure Factor and Spectral Density}
\label{app:Sofk}

For a point configuration in a periodic simulation box $\mathfrak{F}$ in $\R^d$, which consists of $N$ points located at $\vect{r}_1,\ldots, \vect{r}_N$, its structure factor can be written as \cite{hansen_theory_1990, torquato_hyperuniform_2018}
\begin{align} \label{eq:Sk}
    S(\vect{k}) 
    = \frac{1}{\abs{V_\mathfrak{F}}} {\abs{
        \sum_{j=1}^N e^{-i \vect{k}\cdot \vect{r}_j} }^2
    },
\end{align}
where $\abs{V_\mathfrak{F}}$ is the volume of $\mathfrak{F}$.
Considering a packing in the same simulation box, which consists of $N$ spheres of radii $a_1,\ldots,a_N$ and centers $\vect{r}_1,\ldots, \vect{r}_N$, one can evaluate its spectral density as \cite{torquato_hyperuniformity_2016, torquato_perspective_2018}
\begin{align}
    \tilde{\chi}_{_V}(\vect{k}) 
    = \frac{1}{\abs{V_\mathfrak{F}}} {\abs{
        \sum_{j=1}^N \fn{\tilde{m}}{k;a_j} e^{-i \vect{k}\cdot \vect{r}_j} - \phi_2 \int_\mathfrak{F} \dd{\vect{r}} e^{-i\vect{k}\cdot\vect{r}} }^2
    },  \label{eq:spd-packing}
\end{align}
where $\fn{\tilde{m}}{k; a} \equiv \qty(2\pi a/k)^{d/2} \fn{J_{d/2}}{ka}$, and $\phi_2$ denotes the packing fraction.
When all particles have the same radius $a$, Eq. \eqref{eq:spd-packing} is simplified as Eq. \eqref{eq:chi_V-S}.

\section{Details for Model Microstructures}
\label{app:models}

Here, we provide analytic expressions and details in numerical simulations for the five models of two-phase media outlined in Sec. \ref{sec:models}.
Table \ref{tab:summary-model} summarizes their structural features and applicability of predictive formulas for $\ell_s$.
All analytic expressions given here can be evaluated using the Python scripts provided in the \SI.

\subsection{Debye Random Media \label{app:Debye}}

For any space dimension $d$, the specific surface $s$ of this model is expressed as\cite{torquato_random_2002}
\begin{equation} \label{eq:s-Debye}
s= \frac{\phi_1\,\phi_2}{\beta(d)\,a},
\end{equation}
where $\beta(d)$ is given in Eq. \eqref{eq:auto-small-r}.
The spectral density is expressed as \cite{torquato_predicting_2020}
\begin{equation}  \label{eq:chik-Debye}
{\tilde \chi}_{_V}(k) = \frac{ \phi_1\phi_2\, c_d\, a^d}{[1+(ka)^2]^{(d+1)/2}},
\end{equation}
where $c_d=2^d \pi^{(d-1)/2} \Gamma((d+1)/2)$, which is used in Fig. \ref{fig:Sk}. 

\subsection{Equilibrium Packings
\label{app:equilibrium}}

For 2D case, the structure factor is expressed as 
\begin{align} 
S(k) = \frac{1}{1-\rho 2\pi\int_{0}^{1} c(x,\phi_2) J_{0}(2kax)x\dd{x}},
\end{align}
where the direct-correlation function $c(x)$ can be approximated as\cite{guo_theoretical_2006} 
\begin{align} \label{eq:cofr--Guo-Riebel}
&c(x,\phi_2) = \Theta(1-x) \left[ -\frac{1-q \phi_2^2}{(1-2\phi_2 + q\phi_2^2)^2} \right] \Big\{   1-A^2 \phi_2 
\nonumber \\
&+ A^2 \phi_2 \frac{2}{\pi} \qty[ \arccos(\frac{x}{A}) - \frac{x}{A} (1-x^2/A^2)^{1/2} ]
\Big\},
\end{align}
where $x\equiv r/(2a)$, $q\equiv(4 \sqrt{ 3 } \pi - 12)/\pi^2$, and $A(\phi_2)=0.3699 \phi_2^4 - 1.2511 \phi_2^3 + 2.0199 \phi_2^2 - 2.2373 \phi_2 + 2.1$ is a numerical parameter obtained by a polynomial fitting.
For 3D case, the Percus-Yevick solution gives the following expression for $\fn{S}{k}$ \cite{torquato_random_2002}:
\begin{align} \label{eq:3D-Sk}
\fn{S}{k} &= 
\Big(1-\rho \frac{16 \pi  a^3 }{q^6} 
\Big\{\big[24 a_1 \phi_2 - 12 (a_1 + 2 a_2) \phi_2 q^2 
    \nonumber \\
&+ (12 a_2 \phi_2 + 2 a_1 + a_2\phi_2) q^4] \cos(q) 
	\nonumber \\
&+ [24 a_1 \phi_2 q - 2 (a_1 + 2 a_1 \phi_2 + 12 a_2 \phi_2) q^3\big] \sin(q)
    \nonumber \\
&	-24 \phi_2 (a_1 - a_2 q^2) \Big\}\Big)^{-1},
\end{align}
where $q = 2ka$, $a_1 = (1+2\phi_2)^2/(1-\phi_2)^4$, and $a_2 = -(1+0.5\phi_2)^2 /(1-\phi_2)^4$.

\subsection{$g_2$-Invariant Sphere Packing \label{app:g2-inv}}

At the terminal packing fraction, the structure factor of this model is analytically expressed as
\begin{align} \label{eq:step-Sofk-terminal}
S(k) = 1 - \Gamma(1+d/2) \frac{J_{d/2}(kD)}{(kD/2)^{d/2}},
\end{align} 
indicating hyperuniformity with $S(k)\sim k^2$ for small $k$.\cite{torquato_local_2003}

We numerically generate 100 distinct 2D configurations of this hyperuniform model with $N=100$ to conduct FDTD simulations.
These configurations are generated via the Monte Carlo method using the long-ranged effective pair potential:\cite{wang_realizability_2022} 
\begin{align}
    v(r) = 
    \begin{cases}
        +\infty, &  r\leq 1\\
        \epsilon_2 \exp(-r/\sigma_2^{(1)}) \cos(r/\sigma_2^{(2)}+\theta_2) 
        \nonumber \\
        \quad -\epsilon_1 \log(r), & r>1,
    \end{cases}
\end{align}
where the length scale is taken to be unit particle diameter (i.e., $D=1$), $\epsilon_1=4.0$, $\epsilon_2=6.774$, $\sigma_2^{(1)}=0.4250$, $\sigma_2^{(2)}=0.4834$, and $\theta_2 = 1.776$. 

\begin{figure*}[h!t]
    \includegraphics[width=0.45\textwidth]{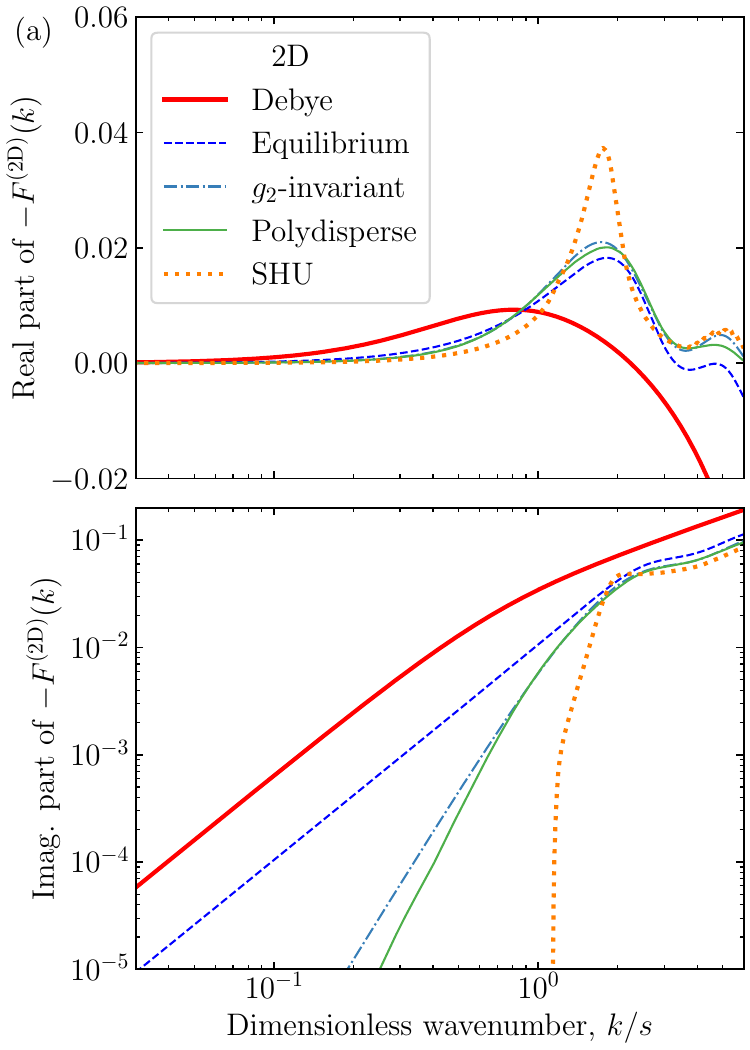}        
    \includegraphics[width=0.45\textwidth]{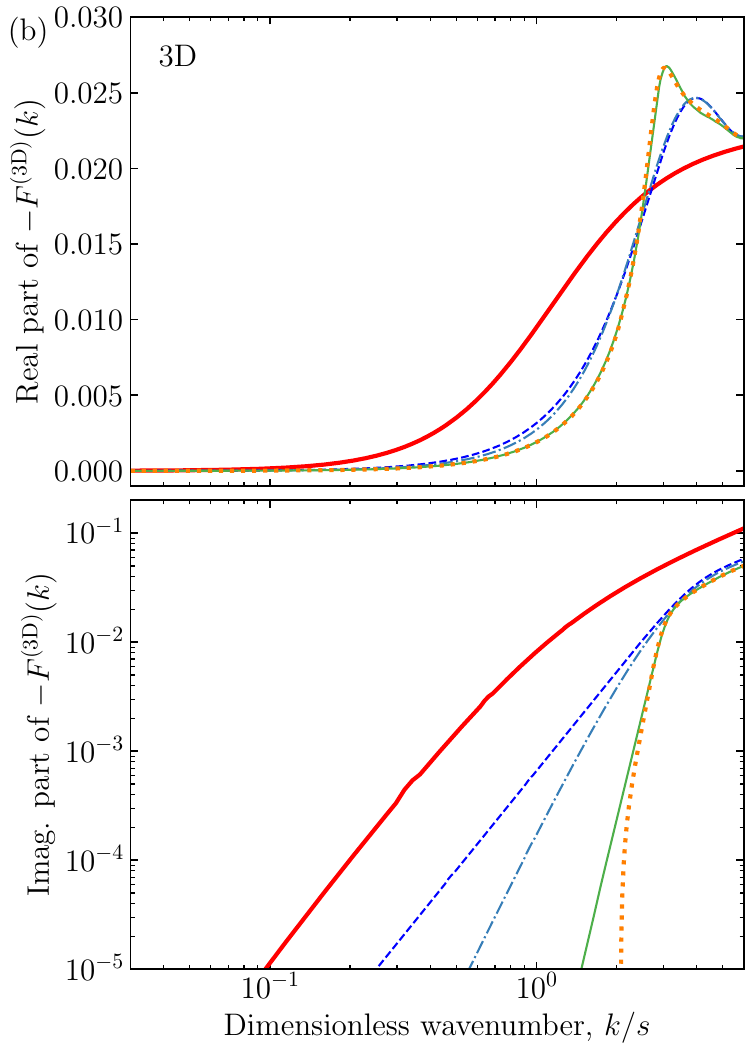}        
    \caption{Negative values of the nonlocal attenuation function, $-F^{\mathrm{(dD)}}(k)$, versus dimensionless wavenumber $k/s$ for the five models of 2D media with $\phi_2=0.25$ (a) and 3D media with $\phi_2=0.125$ (b).
    The top panels show semi-log plots of the real parts of $-\F{d}{k}$ for $d=2,3$, respectively.
    The bottom panels show log-log plots of the imaginary parts of $-\F{d}{k}$ for $d=2,3$, respectively.
    \label{fig:F}
    }
\end{figure*}

\subsection{Hyperuniform Polydisperse Sphere Packings} \label{app:NSHU}

Specifically, the progenitor point patterns are the centers of 2D and 3D equilibrium packings of identical spheres with packing fraction $\phi_b$ that are created by the Monte Carlo method.
One begins with the Voronoi tessellation\cite{torquato_random_2002} of these progenitor point patterns.
We then move the particle center in a Voronoi cell to its centroid and rescale the particle such that the packing fraction inside this cell is identical to a prescribed value $\phi_2<1$.
The same process is repeated over all cells.
The final packing fraction is $\phi_2 = \sum_{j=1}^N \fn{v_1}{a_j}/V_\mathfrak{F} = \rho \fn{v_1}{a}$, where $\rho$ is the number density of particle centers and $a$ represents the mean sphere radius.
The simulation parameters are $\phi_b=0.30$ and $N=100$ for $d=2$, and $\phi_b=0.45$ and $N=100000$ for $d=3$.
Among these 2D packings, we choose 100 distinct ones to conduct FDTD simulations.

\subsection{SHU Sphere Packings}\label{app:SHU}

We numerically generate 2D and 3D SHU sphere packings in the following two-step procedure.
First, we generate point configurations consisting of $N$ particles in a fundamental cell $\mathfrak{F}$ under periodic boundary conditions via the collective-coordinate optimization technique \cite{uche_constraints_2004, batten_classical_2008, zhang_ground_2015}, which finds numerically the ground-state configurations of the following potential energy;\cite{torquato_existence_2025,kim_ultradense_2025}
\begin{equation*}\label{eq:CC_potential}
\fn{\Phi}{\vect{r}^N} =\frac{1}{V_\mathfrak{F}} \sum_{\vect{k}} \fn{\tilde{v}}{\vect{k}}\fn{S}{\vect{k}} +  \sum_{i <j} \fn{u}{r_{ij}},
\end{equation*}
where $V_\mathfrak{F}$ is the volume of $\mathfrak{F}$, $\fn{\tilde{v}}{\vect{k}}=\fn{\Theta}{K-\abs{\vect{k}}}$, $\Theta(x)$ (equal to 1 for $x>0$ and zero otherwise) is the Heaviside step function, $\fn{u}{r}=(1-r/\sigma)^2\fn{\Theta}{\sigma-r}$.
We take ground-state configurations with $\Phi<10^{-19}$ that are still disordered, stealthy, and hyperuniform, and their nearest-neighbor distances are larger than the length scale $\sigma$.
The parameters are $\chi=0.35$, $\phi_c[\equiv\rho v_1(\sigma/2)]=0.60$, $N=100$ for $d=2$, and $\chi=0.35$, $\phi_c=0.50$, $N=4000$ for $d=3$.
Finally, we decorate each point of a ground-state configuration with an identical sphere of radius $a$.

\section{Nonlocal Attenuation Functions of 2D and 3D Models}
\label{app:3D_F}

Here, we present plots of the nonlocal attenuation function $\F{d}{k}$ for $d=2,3$ for the five model microstructures considered in this work.
This complex-valued quantity is crucial for understanding the effective wave behaviors due to correlated disorder in microstructures and for computing $\ell_s$, akin to the self-energy $\Sigma$ in Eq. \eqref{eq:self-energy}.

We begin with the 2D cases with $\phi_2=0.25$; see Fig. \ref{fig:F}(a).
The real part of $\F{2}{k}$, directly related to the effective phase speed of light, is insensitive to the microstructure of a medium, especially in particulate media, as shown in Fig. \ref{fig:F}(a).
Furthermore, $\mathrm{Re}[\F{2}{k}]\sim -k^2$ in the small-$k$ regime applies to all models.\cite{torquato_nonlocal_2021}
In contrast, the imaginary part of $\F{2}{k}$, which indicates attenuation due to scattering, can vary significantly with the microstructure of a medium; see Fig. \ref{fig:F}(a).
The qualitative behaviors of $\F{d}{k}$ for $d=3$ with $\phi_2=0.125$ are similar to those for $d=2$; see Fig. \ref{fig:F}(b).
Thus, in the long-wavelength regime, the imaginary parts for the five models considered here in dimensions ($d=2,3$) can be summarized as follows:
\begin{align} \label{eq:ImF}
    \abs{\mathrm{Im}[\F{d}{k}]} 
    \sim  
    \begin{cases}
        k^{d},   & \text{Debye, Equilibrium}, \\
        k^{d+2}, & \text{$g_2$-invariant},\\
        k^{d+4}, & \text{Polydisperse},\\
        0,       & \text{SHU},\\
    \end{cases}
    \quad k \ll 1.
\end{align}
Note that Eq. \eqref{eq:ImF} is also valid for $d=1$.
Importantly, this observation implies that only hyperuniform media can exhibit different scaling behaviors in $\ell_s$.

\section{FDTD Simulation Setup}
\label{app:FDTD}
\begin{figure*}[t]
\includegraphics[width=0.8\textwidth]{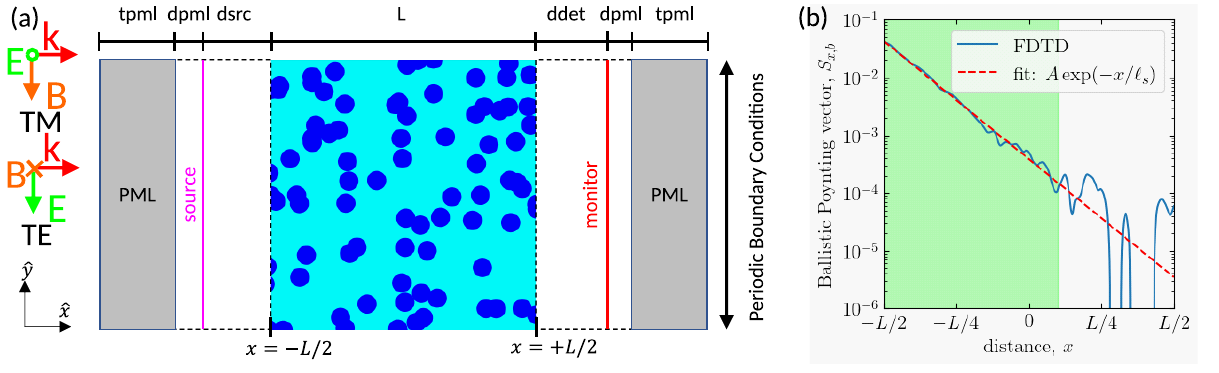}
\caption{ 
 (a) Schematic of the FDTD simulation setup. 
 A linearly polarized Gaussian pulse is generated from a line source (shown in magenta) and is incident on a medium of size $L\times L$ (shown in blue and cyan colors).
 Periodic boundary conditions are applied along the $y$ direction. 
 The perfectly matched layers (PML shown in gray) of thickness \texttt{tpml} are placed at both ends of the simulation box along the $x$ direction.
 The scattering mean free path is estimated from the electric and magnetic fields inside the medium.
 (b) Schematic of the estimation of the scattering mean free path $\ell_s$ from the ballistic Poynting vector $S_{x,b}(x)$ obtained from Panel (a). The fitting range (shown in a green box) changes until its right end reaches $x=L/2$. 
 The value of $\ell_s$ is the fitting parameter that minimizes the least squares error in this varying fitting range.
\label{fig:sim-schem}
}
\end{figure*}

Here, we provide the details of FDTD simulations used to estimate the scattering mean free path $\ell_s$ for the selected 2D model microstructures
described in Sec. \ref{sec:FDTD}.
Since we focus on 2D packing models in our FDTD simulations, all lengths in this section are expressed in units of $\rho^{-1/2}$, where $\rho$ is the number density of particles.

Figure \ref{fig:sim-schem}(a) depicts our simulation setup.
A line source (magenta line) generates a linearly polarized Gaussian pulse propagating in the $x$ direction.
The wavenumber spectrum of this pulse roughly ranges from $k_{\min}$ to $k_{\max}$.
From this source, a two-phase medium of size $L\times L$ (shown in blue and cyan colors) is separated by a distance \texttt{dsrc}($=3.0$).
A monitor (red line) for recording the transmitted electric field is separated from the medium by a distance \texttt{ddet}($=3.0$).
Two perfectly matched layers (PMLs) at both ends of the simulation box absorb any reflected and transmitted waves, and they are separated from the source and monitor by a distance \texttt{dpml}(=0.5).
We apply periodic boundary conditions along the $y$ direction.
The width of the simulation grid is set to be $1/80$ to accurately reproduce the Mie differential scattering cross-section of a single disk up to $k_1\rho^{-1/2}=3.0$.

Simulations are conducted separately for TE and TM polarizations.
For TE polarization, the wavenumber range of the incident pulse is set to be $k_{\min}=0.3$ and $k_{\max}=3.0$, with PML thickness \texttt{tpml}($=1.0 \times 2\pi / k_{\min}$).
In contrast, for TM polarization, which has lower PML performance, the minimum wavenumber increases to $k_{\min}=0.5$. 
Additionally, we thicken PMLs to \texttt{tpml}($=4 \times 2\pi / k_{\min}$).
Each simulation of a medium continues until the transmitted electric field recorded at the monitor decays to $10^{-5}$ of its largest amplitude, or until a maximum time period of $5000/c$ is reached, whichever comes first, where $c$ is the speed of light in vacuum.
Using 20 CPU cores, it takes about 8 h for TM polarization and 1 h for TE polarization.
During simulation, we record the electric field $\vect{E}(\vect{x},t)$ and magnetic field $\vect{H}(\vect{x},t)$ inside the medium as a function of time and position.
Once the simulation ends, we compute the spectra of these fields, denoted by $\vect{E}(\vect{x},k_1)$ and $\vect{H}(\vect{x},k_1)$, at given wavenumber $k_1$ or frequency $f=ck_1/(2\pi)$ by applying the temporal Fourier transform.

As described in Sec. \ref{sec:FDTD}, we calculate the ballistic Poynting vector $S_{x,b}(x)$ at each frequency or wavenumber from ensemble averages of $\vect{E}(\vect{x},f)$ and $\vect{H}(\vect{x},f)$ over 100 realizations of the media.
We then fit $S_{x,b}(x)$ to Eq. \eqref{eq:S_fit} in the range $x\in[-L/2,x_\mathrm{fit}]$, where $x_\mathrm{fit}$ gradually increases from $-L/2$ to $L/2$; see Fig. \ref{fig:sim-schem}(b).
The values of $\ell_s$ are determined by the fitting parameters that minimize the least squares error for different values of $x_\mathrm{fit}$.

\begin{figure*}[h!t]
 \includegraphics[width=0.3\textwidth]{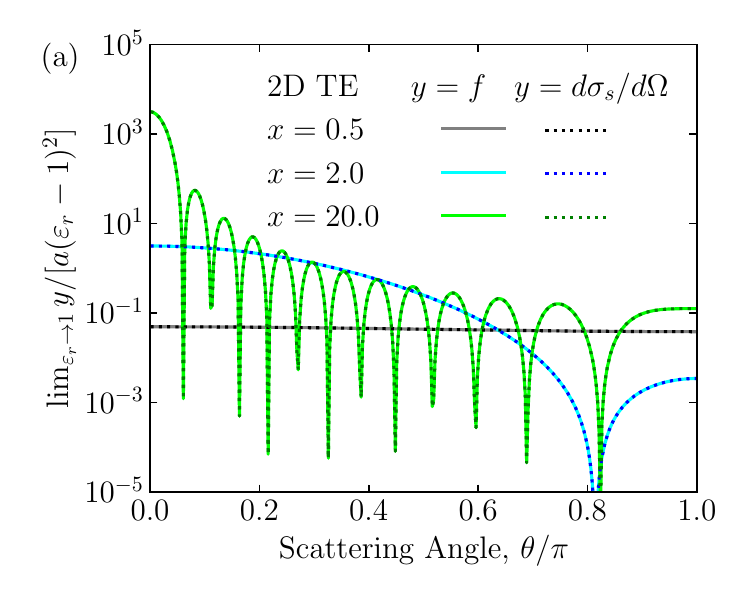}
 \includegraphics[width=0.3\textwidth]{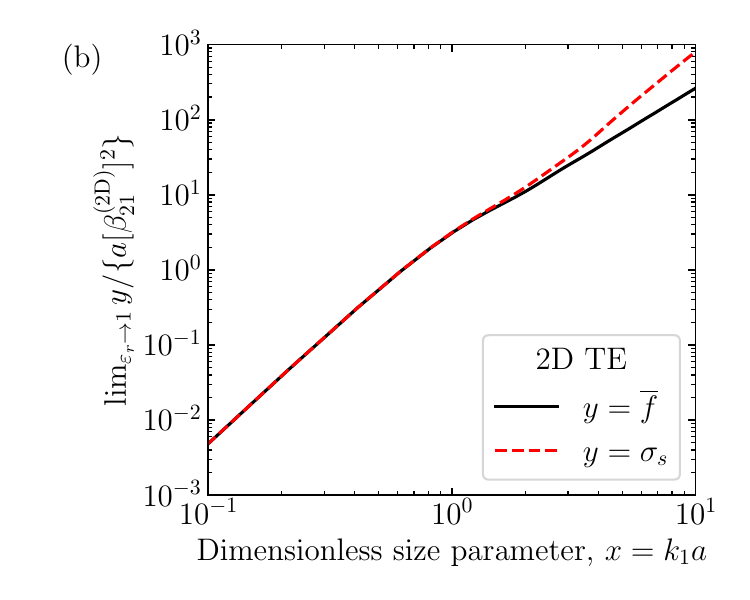}
 \includegraphics[width=0.3\textwidth]{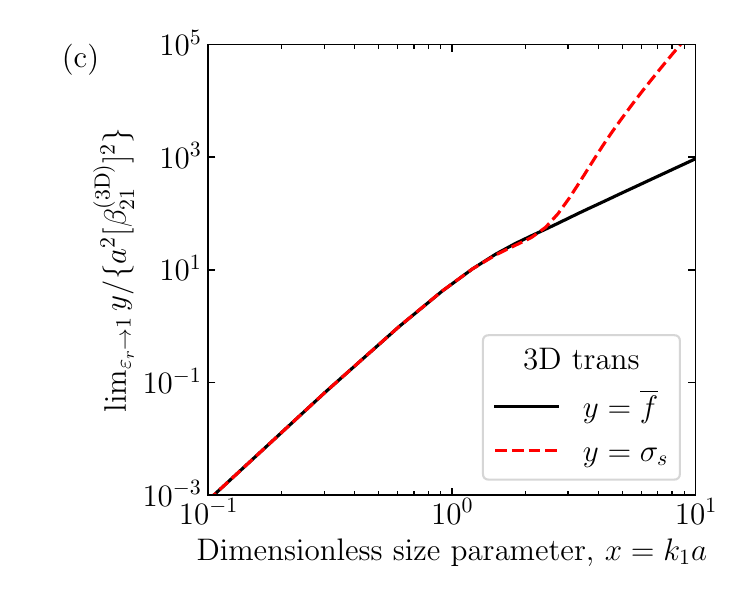}

\caption{Comparisons of the weighted particle form factor and the Mie scattering cross-section of a single particle in the weak-contrast regime, that is, $\varepsilon_r=\varepsilon_2/\varepsilon_1 \to 1$.
(a) 2D TM case: Differential weighted particle form factor $f(k_1,\theta;\varepsilon_r)$, given in Eq. \eqref{eq:form-vs-sigma}, and the leading-order contribution in $(\varepsilon_2-\varepsilon_1)$ of the Mie `differential' scattering cross-section given in Eq. \eqref{eq:diff-2D-TM} are plotted as functions of the scattering angle $\theta$ at three values of $x\equiv k_1 a=0.50,2.0,20.0$.
The $y$ axis is scaled by $a(\varepsilon_2-\varepsilon_1)^2$ so that these two quantities are independent of the contrast ratio.
(b) 2D TM case: Total weighted particle form factor $\overline{f}(k_1;\varepsilon_r)$, given in Eq. \eqref{eq:P-total-2D-TE}, and the leading-order contribution in $\BETA{2}{21}$ of the Mie `total' scattering cross-section given in Eq. \eqref{eq:total-2D-TE} are plotted as functions of $x=k_1a$.
The $y$ axis is scaled by $a[\BETA{2}{21}]^2$ such that both quantities are independent of the contrast ratio.
(c) 3D case: Total weighted particle form factor $\overline{f}(k_1;\varepsilon_2/\varepsilon_1)$, given in Eq. \eqref{eq:P-total-3D}, and the leading-order contribution in $\BETA{3}{21}$ of the Mie `total' scattering cross-section \eqref{eq:total-3D} are plotted as functions of $x$.
The $y$ axis is scaled by $a^2[\BETA{3}{21}]^2$ such that both quantities are independent of the contrast ratio.
 \label{fig:form-vs-sigma}
 }
\end{figure*}

\section{Difference and Similarity Between the Weighted Particle Form Factor and Mie Scattering Cross-Section}
\label{app:FormFactor-Mie}

Here, we numerically compare the Mie scattering cross-section $\dv*{\sigma_s}{\Omega}$ of a single particle and the differential weight particle form factor $f(k_1,\theta;\varepsilon_r)$ given in Eq. \eqref{eq:form-vs-sigma} in the weak-contrast regime, that is, $\varepsilon_r\equiv \varepsilon_2/\varepsilon_1 \to 1$.
The difference and similarity between these two quantities depend on the space dimension and polarization. 
For the 2D TM case, the differential scattering cross-section $\dv*{\sigma_s}{\Omega}$ becomes identical to Eq. \eqref{eq:form-vs-sigma} for any wavenumber $k_1$ and scattering angle $\theta$ in this regime.
In contrast, for the 2D TE and 3D cases, $\dv*{\sigma_s}{\Omega}$ differs from Eq. \eqref{eq:form-vs-sigma}, but their angular integrals become identical in both weak-contrast and intermediate-wavenumber (i.e., $k_1a \lesssim 1.0$) regimes.

We begin with the 2D TM case, in which the leading-order term in $(\varepsilon_2-\varepsilon_1)$ of $\dv*{\sigma_s}{\Omega}$ becomes identical to Eq. \eqref{eq:form-vs-sigma} for any wavenumber $k_1$ and scattering angle $\theta$.
Specifically, in the weak-contrast regime (i.e., $\varepsilon_r=\varepsilon_2/\varepsilon_1 \to 1$), we obtain the following expression from $\dv*{\sigma_s}{\Omega}$ given in Eq. \eqref{eq:diff-sigma-cylinder} for the 2D TM case
\begin{align}
\dv{\sigma_{s}}{\Omega} \approx& 
 \frac{2a}{\pi x} (\varepsilon_2-\varepsilon_1)^2
 \abs{b_{0}' (x) + 2\sum_{n=1}^\infty b_{n}' (x) \cos(n\theta)}^2, \label{eq:diff-2D-TM} 
\end{align}
where $x\equiv k_1a$, and $b_n'(x)$ is the leading-order term in $(\varepsilon_2-\varepsilon_1)$ of the Mie coefficient $b_n$:
\begin{align*}
 b_n'(x) \equiv& 
 \left.\dv{b_n}{(\varepsilon_2-\varepsilon_1)} \right|_{\varepsilon_2=\varepsilon_1}
\\
 =&\frac{\pi x^2}{2} \frac{i}{4}
\qty[J^2_{n-1}(x) + 2\left( 1- \frac{2n^2}{x^2} \right) J_{n}^2(x) + J^2_{n+1}(x) ] . 
\end{align*}
Figure \ref{fig:form-vs-sigma}(a) numerically shows the identity of Eq. \eqref{eq:diff-2D-TM} and Eq. \eqref{eq:form-vs-sigma} by plotting them as a function of $\theta$ for three values of $x=0.5,2.0,20.0$.

For the 2D TE case, from Eq. \eqref{eq:diff-sigma-cylinder}, the leading-order term in $\BETA{2}{21}$ of the differential scattering cross-section $\sigma_s$ is expressed as
\begin{align}
\dv{\sigma_{s}}{\Omega} \approx& 
 \frac{2a}{\pi x} [\BETA{2}{21}]^2
 \abs{a_{0}' (x) + 2\sum_{n=1}^\infty a_{n}' (x) \cos(n\theta)}^2, \label{eq:diff-2D-TE} 
\end{align}
where $a_n'(x)$ is the leading-order term in $\BETA{2}{21}$ of the Mie coefficient $a_n$:
\begin{align*}
a_n'(x) \equiv& 
 \left.\dv{a_n}{\BETA{2}{21}} \right|_{\varepsilon_2=\varepsilon_1}
=\frac{\pi x^2}{2}
\Big[J_{n-1}^2(x) -\frac{2(n-1)}{x} J_n(x) J_{n-1}(x)
\nonumber \\
&+\left(1 -\frac{2 n^2 }{x^2}\right) J_n^2(x) 
\Big]. 
\end{align*}
Unlike the 2D TM case, Eq. \eqref{eq:diff-2D-TE} is not identical to $f(k_1,\theta;\varepsilon_r)$ given in Eq. \eqref{eq:form-vs-sigma} for the 2D TE case.
However, the leading-order term in $\BETA{2}{21}$ of the total scattering cross-section $\sigma_s$, expressed as
\begin{align}  \label{eq:total-2D-TE}
 \sigma_s \equiv \int_0^{2\pi} 
 \dv{\sigma_{s}}{\Omega}\dd{\theta} 
 \approx& 
 \frac{4a}{x} [\BETA{2}{21}]^2
 \sum_{n=0}^\infty
 \abs{a_{n}' (x) }^2,
\end{align}
is approximately the same as the total weighted particle form factor [i.e., the angular integral of Eq. \eqref{eq:form-vs-sigma} for the 2D TE case], given by 
\begin{align}  \label{eq:P-total-2D-TE}
 \overline{f}(k_1;\varepsilon_2/\varepsilon_1) \approx& \frac{[\BETA{2}{21}]^2}{4\pi} {k_1}^{3}
 \nonumber \\
  &\times 
 \int_0^{2\pi} 
 P(2k_1 \sin(\theta/2); a) \dd{\theta},
\end{align}
for intermediate wavenumbers $k_1a \lesssim 1.0$, as shown in Fig. \ref{fig:form-vs-sigma}(b).

Finally, for 3D case, the leading-order term in $\BETA{3}{21}$ of the differential scattering cross-section $\dv*{\sigma_s}{\Omega}$ is expressed from $f(k_1,\theta;\varepsilon_r)$ given in Eq. \eqref{eq:diff-sigma-sphere} as
\begin{align}
&\dv{\sigma_{s}}{\Omega} 
\approx
\frac{[\BETA{3}{21}]^2}{2{k_{1}}^2} 
\nonumber \\
&\times \Bigg[\abs{\sum_{n=1}^\infty \frac{2n+1}{n(n+1)} [a_{n}'(x)\fn{\pi_{n}}{\cos \theta} + b_{n}'(x)\fn{\tau_{n}}{\cos \theta}] }^2 \nonumber \\
&+ \abs{\sum_{n=1}^\infty \frac{2n+1}{n(n+1)} [b_{n}'(x)\fn{\pi_{n}}{\cos \theta} + a_{n}'(x)\fn{\tau_{n}}{\cos \theta}] }^2 \Bigg]
\label{eq:diff-3D} ,
\end{align}
where 
\begin{align*}
a_{n}'(x) =& \frac{3}{2} ix [\left(2 n+x^2+2\right) j_n^2(x)+x^2 j_{n+1}^2(x) 
\nonumber \\
&\quad -(2 n+3) x j_{n+1}(x) j_n(x)],\\
b_{n}'(x) =& \frac{3}{2} i  x^2 [x j_n^2(x)+x j_{n+1}^2(x)
\nonumber \\
&\quad -(2 n+1) j_{n+1}(x) j_n(x)].
\end{align*}
are the leading-order terms in $\BETA{3}{21}$ of the Mie coefficients $a_n$ and $b_n$ for a sphere, respectively.
Similar to the 2D TE case, Eq. \eqref{eq:diff-3D} is not identical to Eq. \eqref{eq:form-vs-sigma} for $d=3$.
Instead, their angular integrals over the entire solid angle, given respectively by 
\begin{align}
\sigma_s \equiv& 
\int_0^{\pi} \dd{\theta} \sin \theta \int_0^{2\pi}\dd{\phi} \dv{\sigma_s}{\Omega}
\nonumber \\
\approx&
\frac{2\pi}{{k_1}^2} [\BETA{3}{21}]^2 \sum_{n=1}^\infty (2n+1)[|a_{n}'(x)|^2+|b_{n}'(x)|^2], \label{eq:total-3D}
\end{align}
and
\begin{align}
  &\overline{f}(k_1;\varepsilon_2/\varepsilon_1) \approx \frac{3[\BETA{3}{21}]^2}{8\pi^2} {k_1}^{4} 
 \nonumber \\
  &\times 
 \int_0^{\pi} \sin\theta \dd{\theta} \int_0^{2\pi} \dd{\phi} 
 P(2k_1 \sin(\theta/2); a) ,\label{eq:P-total-3D}
\end{align}
are approximately the same for intermediate wavenumbers $k_1a \lesssim 1.0$, as shown in Fig. \ref{fig:form-vs-sigma}(c).

% Bibliography embedded for the self-contained arXiv submission.

\end{document}